\documentclass[twocolumn]{aastex631}
\shorttitle{AASTeX v6.31 Sample article}
\shortauthors{Pascucci et al.}

\newcommand{\CI}{[C\,{\scriptsize I}]\,1-0}
\newcommand{\ci}{[C\,{\scriptsize I}]}

\graphicspath{{./}{figures/}}

\begin{document}

\title{Large Myr-old Disks are Not Severely Depleted of gas-phase CO or carbon}

\correspondingauthor{Ilaria Pascucci}
\email{pascucci@arizona.edu}

\author[0000-0001-7962-1683]{Ilaria Pascucci}
\affiliation{Lunar and Planetary Laboratory, The University of Arizona, Tucson, AZ 85721, USA}

\author[0009-0000-9731-2462]{Bennett N. Skinner}
\affiliation{Lunar and Planetary Laboratory, The University of Arizona, Tucson, AZ 85721, USA}

\author[0000-0003-0777-7392]{Dingshan Deng} 
\affiliation{Lunar and Planetary Laboratory, The University of Arizona, Tucson, AZ 85721, USA}

\author[0000-0003-0522-5789]{Maxime Ruaud}
\affiliation{NASA Ames Research Center, Moffett Field, CA 94035, USA}
\affiliation{Carl Sagan Center, SETI Institute, Mountain View, CA 94035, USA}

\author[0000-0002-3311-5918]{Uma Gorti}
\affiliation{NASA Ames Research Center, Moffett Field, CA 94035, USA}
\affiliation{Carl Sagan Center, SETI Institute, Mountain View, CA 94035, USA}

\author{Kamber R. Schwarz}
\affiliation{Max Planck Institute for Astronomy, K\"{o}nigstuhl 17, D-69117, Heidelberg, Germany}

\author{Edwige Chapillon}
\affiliation{Institut de Radioastronomie Millim\'etrique (IRAM), 300 rue de la Piscine, 38406, Saint-Martin d’H\'eres, France}
\affiliation{Laboratoire d’Astrophysique de Bordeaux, Universit\'e de Bordeaux, CNRS, B18N, All\'ee Geoffroy Saint-Hilaire, 33615 Pessac,
France}

\author[0000-0002-4147-3846]{Miguel Vioque}
\affiliation{Joint ALMA Observatory, Alonso de C\'ordova 3107, Vitacura, Santiago 763-0355, Chile}
\affiliation{National Radio Astronomy Observatory, 520 Edgemont Road, Charlottesville, VA 22903, USA}

\author[0000-0002-1575-680X]{James Miley}
\affiliation{Joint ALMA Observatory, Alonso de C\'ordova 3107, Vitacura, Santiago 763-0355, Chile}
\affiliation{National Astronomical Observatory of Japan, NAOJ Chile Observatory, Los Abedules 3085, Oficina 701, Vitacura, Santiago, Chile}








\begin{abstract}
We present an ACA search for \CI\ emission at 492\,GHz toward large T~Tauri disks (gas radii $\gtrsim 200$\,au) in the $\sim 1-3$\,Myr-old Lupus star-forming region. Combined with ALMA 12-m archival data for IM~Lup, we report \CI\ detections in 6 out of 10 sources, thus doubling the known  detections toward T~Tauri disks.
We also identify four Keplerian double-peaked profiles and demonstrate that \CI\ fluxes correlate with  $^{13}$CO, C$^{18}$O, and $^{12}$CO\,(2-1) fluxes, as well as with the gas disk outer radius measured from the latter transition. These findings are in line with the expectation that atomic carbon traces the disk surface. In addition, we compare the carbon and CO line luminosities of the Lupus \& literature sample with \CI\ detections with predictions from the self-consistent disk thermo-chemical models of \citet{Ruaud2022ApJ...925...49R}. These models adopt ISM carbon and oxygen elemental abundances as input parameters. With the exception of the disk around Sz~98, we find that these models reproduce all available line luminosities and upper limits with gas masses comparable to or higher than the minimum mass solar nebula and gas-to-dust mass ratios $\geq 10$. Thus, we conclude that the majority of large Myr-old disks conform to the simple expectation that they are not significantly depleted in gas, CO, or carbon. 
\end{abstract}



\section{Introduction} \label{sec:intro}
Recent high-resolution images of circumstellar disks around young ($\sim 1-10$\,Myr) stars have revealed a variety of complex structures \citep[e.g.,][]{Andrews2020ARA&A..58..483A,Benisty2022}, some of which point to advanced planet formation. Hence, these disks provide an opportunity to study planet formation in action.  Yet, some of their fundamental properties, such as the gas disk mass and the gas-to-dust mass ratio (hereafter $\Delta_{\rm gd}$) $-$ which determine what planets can form as well as the disk lifetime \citep[e.g.,][]{LeeChiang2016} $-$ remain poorly constrained \citep[e.g.,][]{Miotello2022arXiv220309818M}. 

CO is expected to be the most abundant and easily observed tracer of molecular hydrogen, the main gaseous reservoir in young disks, but recent studies have questioned this expectation, especially for young solar analogues (hereafter, T~Tauri stars). ALMA surveys targeting CO isotopologues 
 \citep[e.g.,][]{Ansdell2016ApJ...828...46A,Long2017ApJ...844...99L}
have reported line fluxes far lower than early theoretical estimates obtained by scaling the dust disk mass with the interstellar medium (ISM) $\Delta_{\rm gd}$ of 100 and using the canonical abundance of CO/H$_2 \approx 10^{-4}$ \citep[e.g.,][]{WilliamsBest2014,Miotello2016A&A...594A..85M}.  The mechanisms proposed to explain the CO under-abundance can be grouped into variants of two scenarios: (i) the gas disk has been dispersed and therefore $\Delta_{\rm gd} \ll 100$  \citep[e.g.,][]{WilliamsBest2014,Miotello2017A&A...599A.113M}, or (ii)  $\Delta_{\rm gd}$ is still high but CO is not a good tracer of the gas mass because of chemical processing that transforms CO into other less easily observable species combined with dynamical processes that sequester CO into  midplane ice and redistribute it inward via pebble drift  \citep[e.g.,][]{Bergin2014FaDi..168...61B,Xu2017ApJ...835..162X,Dodson-Robinson2018ApJ...868L..37D,Krijt2020ApJ...899..134K}.  

It is important to note that these early disk models did not include all the relevant physical and chemical processes for interpreting CO emission lines. For instance,  \citet{Aikawa2002A&A...386..622A}, \citet{Thi2010A&A...518L.125T},  \citet{Favre2013ApJ...776L..38F}, and \citet{WilliamsBest2014} include CO freeze-out and/or photodissociation in varying degrees of sophistication but no isotopologue selective dissociation, see \cite{Miotello2014A&A...572A..96M} instead for its implementation. Conversion of CO into CO$_2$ ice on dust grains was also lacking but later found to be significant at radial snowlines where the temperature is $\lesssim 30$\,K and for cosmic-ray ionization rates $\gtrsim 5 \times 10^{-18}$\,s$^{-1}$ \citep[e.g.,][]{Reboussin2015A&A...579A..82R,Eistrup2016A&A...595A..83E,Schwarz2018ApJ...856...85S,Bosman2018A&A...618A.182B}.


Recently,  \citet[][hereafter RGH22]{Ruaud2022ApJ...925...49R} developed models with a self-consistent gas density and temperature structure coupled with vertical pressure equilibrium and further explored the effect of grain surface chemistry on the vertical location of the CO snowline. They found that photoprocessing of the ice by stellar and interstellar FUV photons dominates at the interface between the molecular layer and the disk midplane and efficiently converts CO into  CO$_2$ ice, shifting the vertical CO snowline higher up and effectively reducing the amount of gas-phase CO \citep[see also][]{RG2019ApJ...885..146R}. These new models adopt ISM-like elemental abundances as input parameters and reproduce optically thin C$^{18}$O line fluxes with gas-to-dust ratios of $\sim 100$ without requiring any other chemical or dynamical processes to reduce CO. As such, RGH22 argue that there is no severe CO depletion and  C$^{18}$O emission is a good tracer of the gas disk mass. This argument is supported by Deng et al.~2023 in press (arXiv:4990967), where a favorable comparison between model predictions and observations is extended to C$^{18}$O velocity and radial profiles.

Alongside carbon monoxide, searches for its dissociation products, especially neutral atomic carbon, have been  carried out to investigate elemental abundance depletions. Atomic carbon forms in a thin region between the CO photodissociation and carbon ionization fronts \citep[][]{TH1985ApJ...291..722T}, hence it is expected to be abundant at the surface of protoplanetary disks. In addition, forbidden \ci\ lines are predicted to be optically thin \cite[e.g.,][]{Kama2016}, hence valuable probes of the elemental carbon abundance in the disk surface.  \cite{Chapillon2010A&A...520A..61C} carried out one of the first searches for atomic carbon focusing on CQ~Tau, a Herbig~Ae star whose disk was found to have a low CO-to-dust ratio \citep{Chapillon2008A&A...488..565C}. 
The comparison of their  \ci{} 1-0 and 2-1 line upper limits with several chemical model predictions indicated a $\Delta_{\rm gd}$ of only a few for this disk, suggesting that it may be at a transition phase between protoplanetary and debris. Deeper searches in the \CI\ line toward more sources led to the first bona fide disk detections, one around an Herbig Ae star and two around T~Tauri stars \citep{Tsukagoshi2015ApJ...802L...7T,Kama2016}. Modeling of these lines and other CO isotopologues led \cite{Kama2016theory} to conclude that HD~100546 is at most moderately depleted in gas-phase carbon while  TW~Hya is depleted by two orders of magnitude compared to the ISM value. More recently, \cite{Sturm2022}  reported four new \CI\ disk-like detections and estimated [C/H] depletion factors of $\sim 150$ in DL~Tau, $\sim 15$ in DO~Tau, and only $\sim 5$ in DR~Tau. Clearly, more detections of atomic carbon are necessary to establish the extent of carbon depletion in  Myr-old disks.

Here, we summarize results from our ALMA \CI\ survey targeting large gaseous disks around T~Tauri stars in the nearby  $\sim 1-3$\,Myr-old Lupus star-forming region \citep{Galli2020}. Sect.~\ref{sect:obs_and_redu} discusses our observational strategy and analysis which, combined with archival data for IM~Lup, led to six new  \CI\ detections. In Sect.~\ref{sec:findings} we demonstrate that the Lupus detections are consistent with disk emission and \CI\ fluxes correlate with literature fluxes from $^{12}$CO, $^{13}$CO, and C$^{18}$O, as expected if \CI\ traces the disk surface. We also discuss the Lupus and the literature sample with \CI\ detections in the context of the RGH22 models (Sect.~\ref{sec:modres}) and already published inferences about CO and [C/H] depletion (Sect.~\ref{sec:complit}). Finally, we provide a summary and outlook in Sect.~\ref{sec:summary}.

\begin{longrotatetable}
\begin{deluxetable*}{ccccccccccccccc}
\tablecaption{Stellar and disk properties relevant to this study \label{tab:sample}}
\tablewidth{0pt}
\tabletypesize{\scriptsize}
\tablehead{
\colhead{ID} & \colhead{Name (Other Name)} & \colhead{2MASS} & \colhead{Lupus} & \colhead{Dist.} & \colhead{$M_*$} & \colhead{Log$\dot{M}_{\rm acc}$}& 
\colhead{$F_{\rm 1.3mm}$} & \colhead{$R_{\rm dust}$} & \colhead{$i$} & \colhead{$F_{\rm C^{18}O}$} & \colhead{$F_{\rm ^{13}CO}$}& \colhead{$F_{\rm ^{12}CO}$} & \colhead{$R_{\rm CO}$}&
\colhead{Ref.} \\
\colhead{} & \colhead{} & \colhead{} & \colhead{sub-group} & \colhead{(pc)} & \colhead{(M$_\odot$)} & \colhead{(M$_\odot$/yr)} & 
\colhead{(mJy)} & \colhead{($''$)} & \colhead{(deg)} & \colhead{(Jy\,km/s)} & \colhead{(Jy\,km/s)} & 
\colhead{(Jy\,km/s)}  & \colhead{($''$)}&
\colhead{} 
}
\startdata
1 & Sz~71 (GW~Lup) & J15464473-3430354 & I & 155.20 & 0.41 & -9.03 & 69.15 & 0.63 & -40.8 & 0.076\tablenotemark{\small{a}}& 0.58\tablenotemark{\small{a}}& 2.175& 1.45& 1,2,3\\ 
2 & RY~Lup & J15592838-4021513 & off-cloud & 158  & 1.27 & -8.05 & 86.11 & 0.89 & 68.0 & 0.765& 2.502&6.615 & 1.67 & 1,2,3,4\\ 
3 & SSTc2dJ160002.4-422216 & J16000236-4222145 & IV & 160.39 & 0.19 & -9.48 & 49.96 & 0.75 & 65.7 & 0.052& 0.976& 2.96&1.77&  1,2,4 \\ 
4 & Sz~133 & J16032939-4140018 & IV& 158 & $\approx$0.7\tablenotemark{\small{b}} & -- & 27.02 & 0.95 & 78.5 & $<$0.054 & 0.282& 2.12 &1.59 & 5,2,4 \\ 
5 & Sz~91 & J16071159-3903475 & III & 159.39 & 0.52 & -9.08 & 9.52 & 0.77 & 51.7 & $<$0.087& 1.097& 2.7 & 2.25  & 1,2,4 \\ 
6 & Sz~98 (HK~Lup,V1279~Sco) & J16082249-3904464 & III & 156.27 & 0.55 &  -7.44 & 103.35 & 0.95 & -47.1 & $<$0.054& $<$0.081& 3.55& 1.79 & 1,2,4 \\ 
7 & SSTc2dJ160830.7-382827 & J16083070-3828268 & III & 158 & 1.27 & -9.2 & 38.76 & 0.91 & -74.0 & 1.454& 2.736& 7.281 & 1.97 & 1,2,4 \\ 
8 & V1094~Sco & J16083617-3923024 & III & 158 & 0.83 & -8.01& 180.0 & 1.67 & -55.4 & 1.0\tablenotemark{\small{a}} & 6.6\tablenotemark{\small{a}}& 29 & 2.19 & 1,6,7,3,4 \\ 
9 & Sz~111 & J16085468-3937431 & III & 158.37 & 0.52 & -9.47& 60.29 & 0.67 & -53.0 & 0.586& 2.187& 5.963 & 2.31 &1,2,4 \\ 
\hline
10 & IM~Lup (Sz~82)\tablenotemark{\small{c}} & J15560921-3756057 & II & 155.82 & 0.72 & -7.85&  205.0 & 1.5 & -48.0 & 1.325& 5.893& 13.7 & 2.6 & 1,2,4 
\enddata
\tablecomments{$F_{C^{18}O}$, $F_{^{13}CO}$, and $F_{\rm ^{12}CO}$ are line fluxes for the 2-1 transition of the respective CO isotopologues. $R_{\rm dust}$ and $R_{\rm CO}$ are the dust and gas radii containing 90\% of the continuum and of the $^{12}$CO(2-1) transition flux.}
\tablenotetext{a}{The reported line fluxes in \citet{Ansdell2018} are significantly different from those in Deng et al. in prep., which rely on deeper ALMA exposures. Therefore, for this study, we have opted to utilize the latter values, which come with an estimated uncertainty of 20\%.}
\tablenotetext{b}{Under luminous source due to edge-on disk, approximate mass  from effective temperature and cluster age.}
\tablenotetext{c}{IM~Lup was not part of our ACA survey because of already available ALMA Band~8 12-m data covering the \CI\ line}
\tablerefs{1. \citet{Galli2020}; 2. \citet{Manara2022}; 3. Deng et al. in prep.; 4. \citet{Ansdell2018}; 5. \citet{Comeron2008}; 6. \citet{Sanchis2021}; 
6. \citet{Alcala2017} }
\end{deluxetable*}
\end{longrotatetable}

\section{Observations and Analysis} \label{sect:obs_and_redu}
Our ACA sample was selected from the ALMA Band~6 Lupus survey \citep{Ansdell2018} to include disks with large gas outer radii as measured from the $^{12}$CO\,(2-1) transition and a broad range of millimeter continuum flux densities. The first criterion was applied to boost the \CI\ detection rate because this line is expected to probe gas as far out as the $^{12}$CO\,(2-1) line, see Sect.~\ref{sec:modres}. The second criterion was applied to investigate disks with a large range of dust (and possibly gas)  masses. IM~Lup, which hosts one of the largest disks in Lupus \citep[e.g.,][]{Cleeves2016ApJ...832..110C}, was not included in our ACA sample because already available ALMA 12-m data cover and detect the \ci{} 1-0 line. Table~\ref{tab:sample} presents the properties of both our ACA sample and IM~Lup that are relevant to this study, including their respective stellar and disk characteristics. When collecting the literature isotopologue fluxes from \citet{Ansdell2016ApJ...828...46A}, we noticed an unrealistic C$^{18}$O upper limit of 0.07\,Jy\,km/s for the large disk of V1094~Sco. van Terwiska priv. comm. commented that this upper limit is unreliable because it was computed over the 0.25$''$ beam size of shallow observations. Thankfully, V1094~Sco, as well as Sz~71, which was undetected in in both $^{13}$CO and C$^{18}$O in \citet{Ansdell2016ApJ...828...46A}, have deeper CO isotopologue exposures through the  ALMA Large Program AGE-PRO (2021.1.00128.L, PI: K. Zhang). These newer observations detect both disks in $^{13}$CO and C$^{18}$O and find that V1094~Sco is $\sim 14 \times$ brighter in C$^{18}$O while Sz~71 is $\sim 7 \times$  brighter in $^{13}$CO than indicated by the \citet{Ansdell2016ApJ...828...46A} upper limits. As such, we adopt the AGE-PRO CO isotoplogue fluxes from Deng et al. in prep. in this study.  

\subsection{Observations}\label{sect:obs}
Our  ACA Band~8 data were acquired between February 2020 and August 2021 as part of the program 2019.1.00927.S (PI: I. Pascucci, ALMA Cycle~7). A scheduling block including all nine sources was repeated 19 times during this time frame; 13 executions passed Q0 quality assurance, enabling further calibration. The  setup included a spectral window (SPW) centered around the \CI\ line at 492.161\,GHz with a total bandwidth of 1\,GHz and 2048 channels (spectral resolution $\sim 0.3$\,km/s) as well as a main continuum SPW centered at 491\,GHz with a  bandwidth of 2\,GHz and 128 channels. The other two SPWs were centered around 480.269\,GHz and 478.633\,GHz for the serendipitous discovery of CH$_3$OH emission and had  2\,GHz bandwidth with 2048 channels. As no CH$_3$OH lines were detected, these two latter SPWs are also used to image the continuum (Sect.~\ref{sect:dataredu}). Requested exposure times were $\sim 1.3$\,h per source to achieve an rms of 0.1\,Jy/beam over twice the spectral resolution in the SPW covering the \CI\ line. Actual exposure times per source varied from 37\,min for Sz~71 to 56\,min for V1094~Sco, see Appendix~\ref{app:ALMAlog} for details.    

The IM~Lup 12-m data were acquired in March 2016 as part of the program 2015.1.01137.S (PI: T. Tsukagoshi, ALMA Cycle~3). The setup included two SPWs  with 2\,GHz bandwidth for the continuum centered at $\sim$\,480 and $\sim$\,478\,GHz and two SPWs with 59\,MHz bandwidth and 240 channels (channel width 0.15\,km/s) to cover the \CI\ and the CS\,(10$-$9) lines. IM~Lup was observed for $\sim 9$\,min with 41 antennas  delivering a synthesized beam of 
$0.37 '' \times 0.32 ''$ and PA of 75$^\circ$.

\subsection{Data reduction and analysis}\label{sect:dataredu}
\paragraph{ACA Sample} 
The ACA data were initially manually calibrated by the North American ALMA Science Center using the CASA pipeline version 6.2.1.7. Using the same pipeline version, we first split off the calibrated data and concatenate the 13 executions for our targets.
Next, we flag the channels where the \CI\ line could be detected ($500-1150$ for SPW 0) and channels where there is strong water vapor absorption ($780-850$, $1250-1490$, and $1770-1870$ in SPW2) and generate a continuum measurement set per target. From this set, we produce a first image per target using the task \texttt{tclean} with Briggs weighting, robust=0.5, and a shallow threshold of 10\,mJy which is $\sim 3$ times the expected rms from the ALMA sensitivity calculator for an exposure of 45\,minutes in Band~8. We also use a mask centered at each target's coordinates (obtained from a \texttt{uvmodelfit} on the continuum measurement set) that is two times the ACA synthesized beam\footnote{At 491\,GHz the primary beam of a 7m ACA antenna is $\sim 21 ''$ while that of a 12m ALMA antenna is $\sim 13 ''$} ($3.2 '' \times 2 ''$, PA of -68$^\circ$) to cover most of the expected \CI\ emission based on the CO emitting radii (see Table~\ref{tab:sample}). All sources are detected in the continuum. The \texttt{no-selfcal} columns in  Table~\ref{tab:res} provide a  first estimate of the peak signal-to-noise (hereafter, S/N) within the mask and of the rms in an annulus from 7$''$ to 8$''$ centered around each target. 

We also performed self-calibration on the continuum for all sources. After experimenting with the \texttt{gaincal} input parameters, we found that the best results were achieved by performing one phase self-calibration combing all spectral windows and scans with an infinite solution interval and by excluding the last execution. This step yielded improved rms, hence peak-to-rms S/N, from factors of $\sim$\,1.5 to $\sim$\,3. 
Using the same parameters, one amplitude self-calibration slightly improved the rms from $\sim$\,3\% up to $\sim$\,35\% depending on the source, except for Sz~133 and Sz~91. As such, amplitude self-calibration was not applied for these two sources. The  continuum \texttt{selfcal} columns in Table~\ref{tab:res} provide the achieved rms and peak-to-rms S/N from the  primary-beam corrected images. Figure~\ref{fig:continuum} in Appendix~\ref{app:Continuum} shows the continuum images and the ellipse obtained by fitting a 2D Gaussian with the \texttt{imfit} command. The only source that is marginally resolved in the continuum is V1094~Sco with  major and minor axes of $4.1 '' \times 2.6 ''$. The source flux density ($F_{\rm 0.6mm}$) and associated uncertainty obtained via \texttt{imfit} are also summarized in Table~\ref{tab:res}.

\begin{deluxetable*}{lcc|ccc|cccc}[h]
\tablecaption{Results on  primary-beam corrected images and datacubes.  \label{tab:res}}
\tablewidth{0pt}
\tablehead{
\colhead{Source} & \multicolumn{2}{c|}{Cont. \texttt{no-selfcal}} & \multicolumn{3}{c|}{Cont. \texttt{selfcal}} 
& \multicolumn{4}{c}{\CI\ \texttt{selfcal}}\\
\colhead{} & \colhead{rms} & \colhead{S/N} & \colhead{rms} & \colhead{S/N} & \colhead{$F_{\rm 0.6mm}$} &
\colhead{rms} & \colhead{$F_{\rm CI}$} & \colhead{$v_{\rm c, lsr}$} & \colhead{$\sigma$}\\
\colhead{} & \colhead{(mJy/beam)} & \colhead{} & \colhead{(mJy/beam)} & \colhead{} & \colhead{(mJy)} &
\colhead{(Jy/beam)} & \colhead{(Jy km/s)} & \colhead{(km/s)} & \colhead{(km/s)} 
}
\startdata
Sz~71 & 4.94 & 46 & 1.80 & 133 &  255.1$\pm$2.3 & 0.13 & $<$1.23 & & \\ 
RY~Lup & 4.73 & 100 &  1.96&  242& 501.4$\pm$2.5 & 0.13 & 1.52$\pm$0.28  & 4.8$\pm$1.2 & 2.8$\pm$0.7 \\
J16000236 & 2.45 & 70& 1.35& 126& 179.8$\pm$2.1 & 0.13 & $<$0.85 & & \\ 
Sz~133 &  2.57 & 42 & 1.75 & 62& 115.8$\pm$2.1 & 0.14 & $<$0.59 & & \\ 
Sz~91 &  2.56 & 32 &  1.98&  42& 93.1$\pm$2.5 & 0.16 & 2.90$\pm$0.19 & 3.6$\pm$0.1 & 1.4$\pm$0.1 \\ 
Sz~98  & 4.31  & 69 & 2.24 & 135 & 337.3$\pm$2.4 & 0.13 & $<$1.77 & & \\ 
J16083070 & 3.76 & 71 & 2.05 & 133  &  281.3$\pm$2.4 & 0.13 & 2.47$\pm$0.12 & 5.2$\pm$0.1 & 2.5$\pm$0.1 \\ 
V1094~Sco & 7.97 & 96 &  2.78& 280 &  1220.6$\pm$8.8 & 0.12 & 4.99$\pm$0.25 & 5.22$\pm$0.08 & 1.26$\pm$0.06 \\ 
Sz~111 & 3.63 & 87 & 1.84 & 173& 339.1$\pm$2.8 & 0.14 & 2.45$\pm$0.13 & 4.18$\pm$0.07 & 1.06$\pm$0.05 \\
\hline
IM~Lup\tablenotemark{\small{a}} & 0.85  & 226 & 0.67 & 306 &  1574.0$\pm$34 & 0.04 & 11.3$\pm$0.5\tablenotemark{\small{b}} & 4.57$\pm$0.07 & 1.5\tablenotemark{\small{b}}\\
\enddata
\tablenotetext{a}{IM~Lup is not part of our ACA survey and results reported here are from archival ALMA 12-m data, see text for more details.}
\tablenotetext{b}{A Gaussian profile is not a good representation for the extracted \CI\ velocity profile of IM~Lup. Hence, the line flux ($F_{\rm CI}$) is obtained from straight integration and $\sigma$ is calculated directly from the FWHM of the spectrum. }
\end{deluxetable*}

\begin{figure}[t]
    \centering
    \includegraphics[width = 0.5\textwidth]{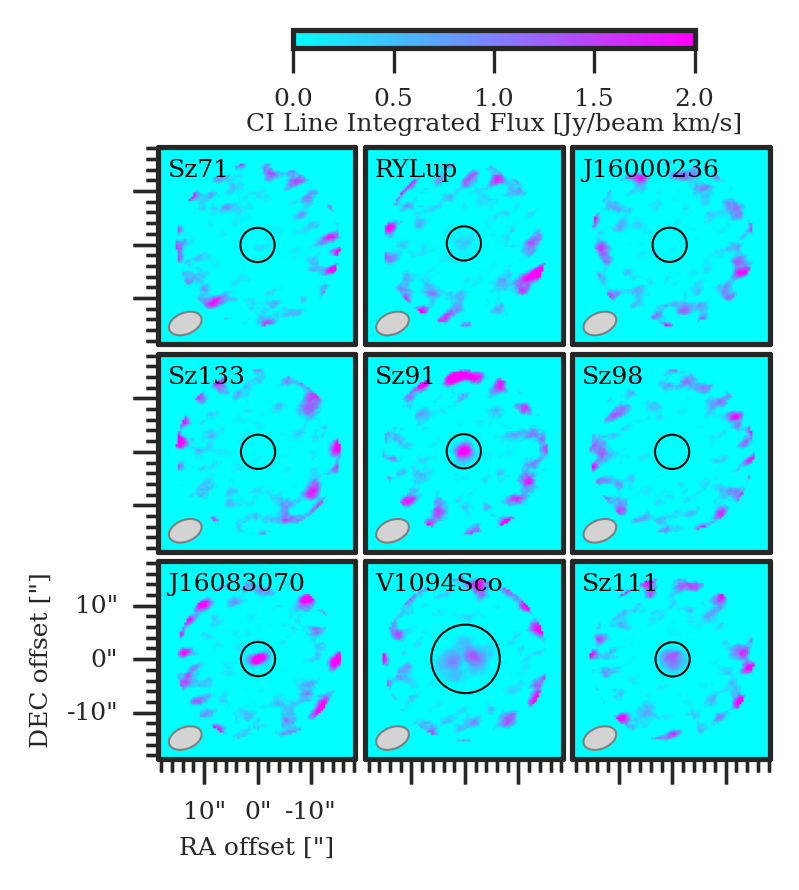}
    \caption{\CI\ moment zero maps obtained with \texttt{bettermoments}  from self-calibrated datacubes. The black circle indicates the extraction region used in \texttt{GoFish} to obtain the spectra in Figure~\ref{fig:CIspectra}. The ACA beam is shown with a grey ellipse in the bottom left of each panel. The extraction radius for V1094~Sco is twice the beam major axis, see text for details.}
    \label{fig:CImom0}
\end{figure}

\begin{figure}[b]
    \centering
    \includegraphics[width = 0.5\textwidth]{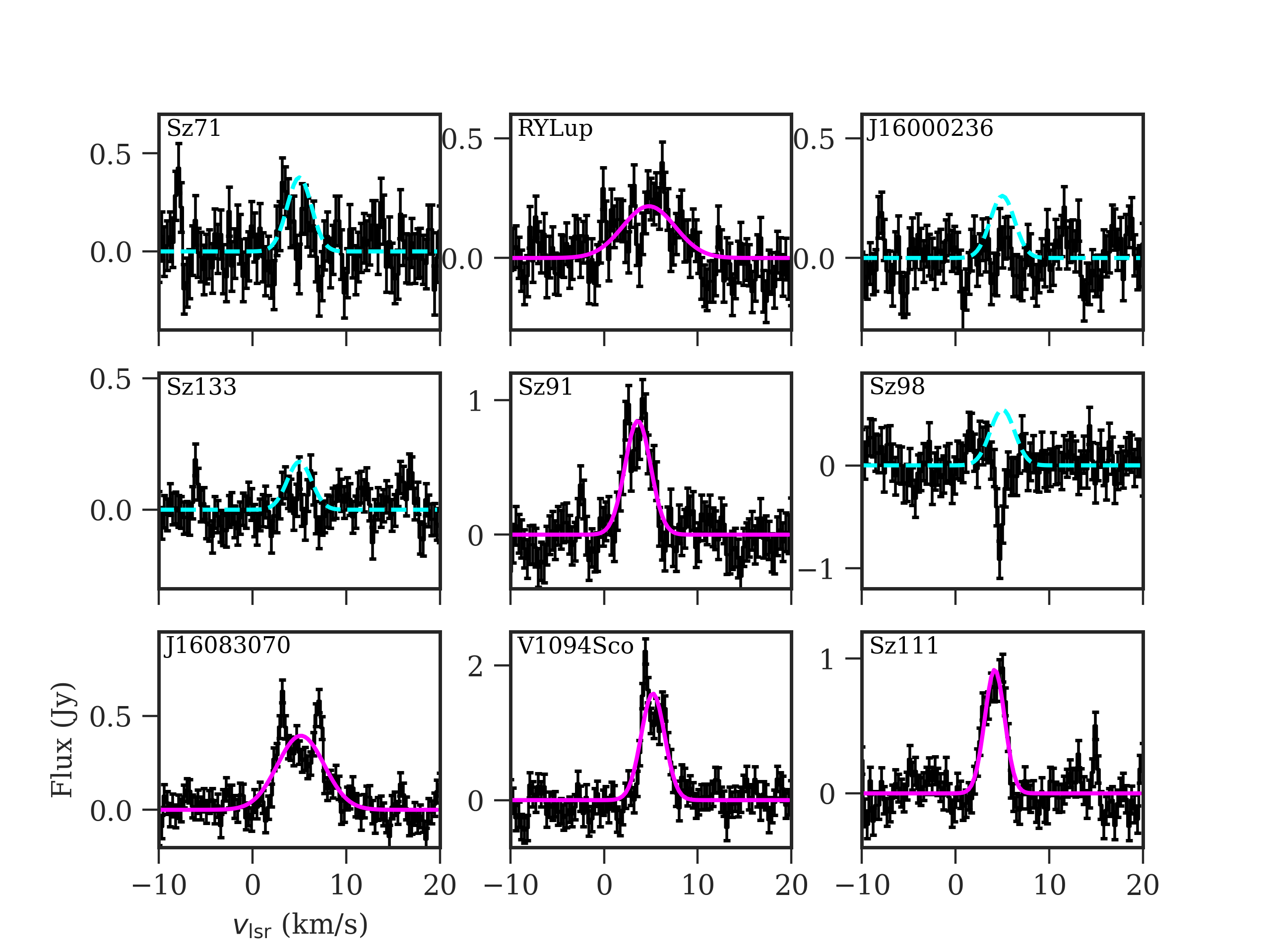}
    \caption{ACA \CI\ non-deprojected spectra (black) extracted with \texttt{GoFish} within the circle shown in Figure~\ref{fig:CImom0}. On top of the spectra we plot the best Gaussian fit when the line is detected (magenta solid line) and the hypothetical 3$\sigma$ upper limit (cyan dashed line) when the line is not detected, see also Table~\ref{tab:res}.
    }
    \label{fig:CIspectra}
\end{figure}

To image the \CI\ line we first performed a continuum subtraction in the spectral window covering the transition with the command \texttt{uvcontsub}. Next, we produced initial datacubes with \texttt{tclean} down to the threshold of 10\,mJy/beam as the initial continuum images. In parallel, we applied to the continuum-subtracted measurement sets the phase and, when available, amplitude solutions obtained on the continuum. We then cleaned these self-calibrated data  down to 3 times the rms of the continuum images. 
To evaluate if the \CI\ emission is detected and the effect of self-calibration, we extract the non-deprojected spectra using \texttt{GoFish} v1.5 \citep{GoFish} with an outer radius equal to the major axis of the beam. We also compute moment zero (integrated intensity) maps with \texttt{bettermoments} \citep{Teague2018zndo...1419754T} with a sigma clipping two times the rms and within  channels corresponding to  velocities where emission is detected in the spectra ($V_{\rm LSR}= 4.5 \pm 5$\,km/s). By comparing these products we find that the rms is essentially unchanged between the no self-calibrated and the self-calibrated datacubes and is slightly larger than the requested one (see Table~\ref{tab:res}). For the sources with a \CI\ detection (RY~Lup, Sz~91, J16083070, V1094~Sco, and Sz~111) the line flux is typically improved but only by $\sim 5-10$\%. This negligible improvement in Band~8 ACA line data after self-calibration has been also noted by \citet{Sturm2022}. Nevertheless, we proceed with the self-calibrated datacubes and fit a 2D Gaussian to the moment zero maps with \CI\ emission (Figure~\ref{fig:CImom0}) to estimate the outermost radius for the extraction of the spectra. We find that for RY~Lup, Sz~91, J16083070, and Sz~111 the \CI\ emission is confined within the ACA primary beam while for V1094~Sco \texttt{imfit} estimates a major and minor axis of 8.1$\times$4.2$''$. To cover most ($> 90$\%) of the emission, we adopt 3.2$''$ as the extraction radius for all sources except for V1094~Sco, for which we use a radius of 6.4$''$. We have checked that for V1094~Sco this radius encompasses all the emission within 2 times the rms in the moment zero map and tested that further increasing the extraction radius  results in a significantly larger increase in the noise than in the line flux. 
The non-deprojected extracted spectra are shown in Figure~\ref{fig:CIspectra}. Of the sources with a \CI\ detection, the spectra from Sz~91, J16083070, and V1094~Sco show a double-peaked Keplerian profile demonstrating that this line traces disk emission.  A single Gaussian is a good representation for all the profiles (see Figure~\ref{fig:CIspectra}) and we have verified that for all sources, including Sz~91, J16083070, and V1094~Sco, a straight integration under the \CI\ line gives the same  flux as the one obtained from the Gaussian fit within the uncertainties quoted in Table~\ref{tab:res}, see $F_{\rm CI}$ column. These uncertainties are obtained in a Monte Carlo fashion. First, we generate 1,000 spectra per source  by randomizing the flux density at each velocity bin from a normal distribution with a  standard deviation equal to the rms outside the line. Next,  we fit a Gaussian to each randomly-generated spectrum and take as uncertainty the standard deviation of the 1,000 Gaussian fluxes. In  case of non-detections, we fit a first-order polynomial between -30 and -10\,km/s and calculate the rms as the standard deviation of the data minus the best fit. Table~\ref{tab:res}  reports a 3$\sigma$ upper limit obtained from this rms and a Gaussian line profile with a line width of 1.3\,km/s (the median value of the \CI\ detections) is shown in Figure~\ref{fig:CIspectra} with a cyan dashed line.

\begin{figure}[h]
    \centering
    \includegraphics[width = 0.5\textwidth]{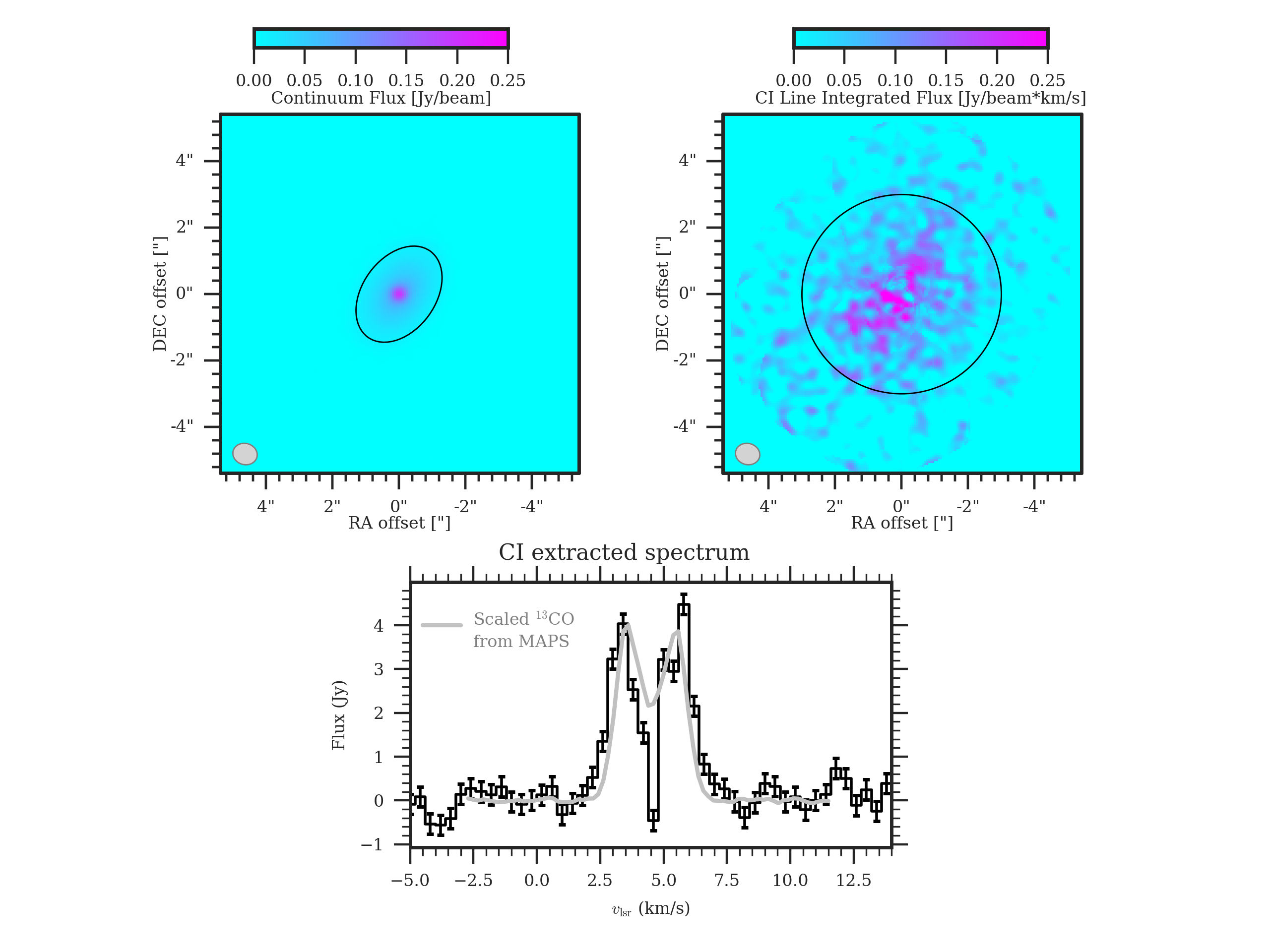}
    \caption{Results for IM~Lup obtained from ALMA 12-m archival data. Upper left panel: continuum emission with best fit 2D Gaussian from \texttt{imfit} (black ellipse). Upper right panel: \CI\ moment zero map with the extraction region for the spectrum (black circle). Lower panel: non-deprojected \CI\ spectrum  (black) extracted with \texttt{GoFish}. The spectral resolution has been degraded to 0.4\,km/s. We also superimpose in grey the scaled $^{13}$CO spectrum of IM~Lup extracted with \texttt{GoFish} from the MAPS fully calibrated datacubes \citep{Oberg2021ApJS..257....1O}. Note the similarity of the two profiles. }
    \label{fig:IMLup}
\end{figure}

\paragraph{IM~Lup}
We retrieved ALMA 12-m archival data that were first manually calibrated by the ALMA NAOJ with the CASA pipeline version 4.6.0. We use the more recent 6.5.0 version for subsequent processing of the only execution block available for this observation. First, we flagged all the channels where the \CI\ line could be detected (51-188) as well as additional channels with apparent emission lines and generate a continuum measurement set. Next, we self-calibrate the continuum combining spectral windows and scans to improve the  S/N ratio. We performed three iterations of phase-only self-calibrations (intervals of 360, 240, and 160\,s) and then one amplitude self-calibration. The reference antenna (DV16) was chosen from the log based on its data quality and position in the array. The self-calibrated continuum visibility was then imaged with the \texttt{tclean} task using a Briggs robust parameter of 0.5 and an elliptical mask  ($2.3\arcsec \times 1.7\arcsec$ with PA = $145^{\circ}$) encompassing the emission. Table~\ref{tab:res} summarizes the improvement in the continuum S/N. We then apply the  calibration tables to the original unflagged and spectrally unaveraged visibilities and split the \CI\ spectral window for the following line imaging. We subtract the continuum using the task \texttt{uvcont-sub} and produce a preliminary \CI\ datacube using the \texttt{tclean} task with Briggs robust = 0.5 and an elliptical mask that encloses the emission. In the preliminary line datacube, the \CI\ emission shows a clear Keplerian pattern. As such, we construct a Keplerian mask to CLEAN again the continuum-subtracted visibilities. The Keplerian mask uses the disk inclination and position angle from a Gaussian fit to the continuum ($49^{\circ}$ and $145^{\circ}$, respectively), the mass of IM~Lup ($0.72\,M_\odot$, see Table~\ref{tab:sample}), and an outer radius that is large enough to include all the emission seen in the preliminary datacube. The CLEANed spectral line cube achieves better image quality with smaller rms (Table~\ref{tab:res}) compared to the pipeline-generated non self-calibrated data. The self-calibrated continuum and \CI\ moment zero maps are shown in the upper panels of Figure~\ref{fig:IMLup}.

To characterize the continuum and \CI\ emission we apply similar steps to those described for our ACA sample. The continuum flux is obtained by fitting a 2D Gaussian with the \texttt{imfit} command. The \CI\ spectrum is extracted with \texttt{GoFish} using a maximum radius of 3$''$ which maximizes the flux and encompasses the emission from the moment zero map. Because a Gaussian profile is not a good representation of the extracted spectrum  (see lower panel of Figure~\ref{fig:IMLup}), the flux reported in Table~\ref{tab:res} is from direct integration under the emission  line and the line width is also measured directly on the extracted profile. We also checked that the flux obtained via integration under the line is the same within the quoted uncertainty as that obtained from the \texttt{GoFish} stacked deprojected spectrum using a Gaussian fit.

\section{Results and Discussion} \label{sec:findings}
All of the 10 Lupus disks investigated here are firmly detected in Band~8 in the continuum with S/N ranging from 42 to 280 (see Table~\ref{tab:res}). The dust continuum emission is confined within the large ACA  beam ($3.2 '' \times 2 ''$) for all sources except for V1094~Sco for which we report a marginal extension (2D Gaussian of $4.1 '' \times 2.6 ''$). This is in line with previous  1\,mm ALMA observations that find its continuum emission extending out to 300\,au from the star, a radial extension only comparable to IM~Lup and $\sim 5$ times larger than other disks in Lupus \citep{vanTerwisga2018A&A...616A..88V}. The \CI\ line is detected in 6 out of 10 disks with integrated fluxes that range from $\sim 5$ (RY~Lup) to $\sim 23$ (IM~Lup) times the reported uncertainties, see Table~\ref{tab:res}.   First, we discuss empirical evidence that the \CI\ line in our Lupus sample traces the gaseous disk surface (Sect.~\ref{sec:CIindisks}). Next, we compare the \ci{} and CO isotopologue luminosities to the RGH22 theoretical predictions and find no need to invoke significant CO or carbon depletion for the large disks discussed in this paper (Sect.~\ref{sec:modres}). Finally, we examine these findings in the context of published gas-to-dust  and [C/H] ratios (Sect.~\ref{sec:complit}).

\subsection{\CI\ emission as a probe of the gas disk surface} \label{sec:CIindisks}
The first piece of evidence favoring disk emission for the \CI\ line comes from the  velocity centroids and profiles which are resolved even at the ACA spectral resolution of $\sim 0.3$\,km/s. The \CI\ line centroid ($v_{\rm c, lsr}$ in Table~\ref{tab:res}) of each source falls within one standard deviation of the median of the stars in its Lupus sub-group (see Table~\ref{tab:sample} and \citealt{Galli2020}).
For RY~Lup, Sz~111, and IM~Lup, whose stellar radial velocities have been precisely measured via high-resolution optical spectra \citep{Fang2018,Banzatti2019ApJ...870...76B}, the \CI\ centroids are within $2\sigma$ of the reported values.
None of the Lupus \CI\ profiles exhibit signs of outflowing or infalling material, unlike the profiles of FM~Cha and WW~Cha observed in \citet{Sturm2022}. This difference is likely due to the fact that the Lupus sources have a lower visual extinction ($A_V < 2$) and are more evolved than those selected by \citet{Sturm2022}.
The median \CI\ FWHM of 3\,km/s suggests broadening beyond thermal effects. If we consider the temperature corresponding to the upper energy level of the transition (23.6\,K), the line width would only be 0.3\,km/s. However, a FWHM of 3\,km/s is consistent with Keplerian broadening around a solar-mass star for a characteristic emitting radius of 100\,au, which is the expected radius according to gas disk models (e.g., Fig.~4 in \citealt{Kama2016}). In fact, the \CI\ profiles from IM~Lup and J16083070, and to a lesser extent, Sz~91 and V1094~Sco, show double-peaked profiles as expected from gas in a Keplerian disk.  In the case of IM~Lup, we have also the advantage of deeper observations of its CO isotopologues through the ALMA MAPS program \citep{Oberg2021ApJS..257....1O}. The lower panel of Figure~\ref{fig:IMLup}  demonstrates the remarkable similarity between the velocity profile from the $^{13}$CO\,(2-1) line, which is tracing the disk surface \citep[e.g.,][]{Law2021ApJS..257....4L}, and the \CI\ line. This comparison suggests that the \CI\ line probes the surface of a Keplerian disk. 

\begin{figure*}[h]
    \centering
    \includegraphics[width = \textwidth]{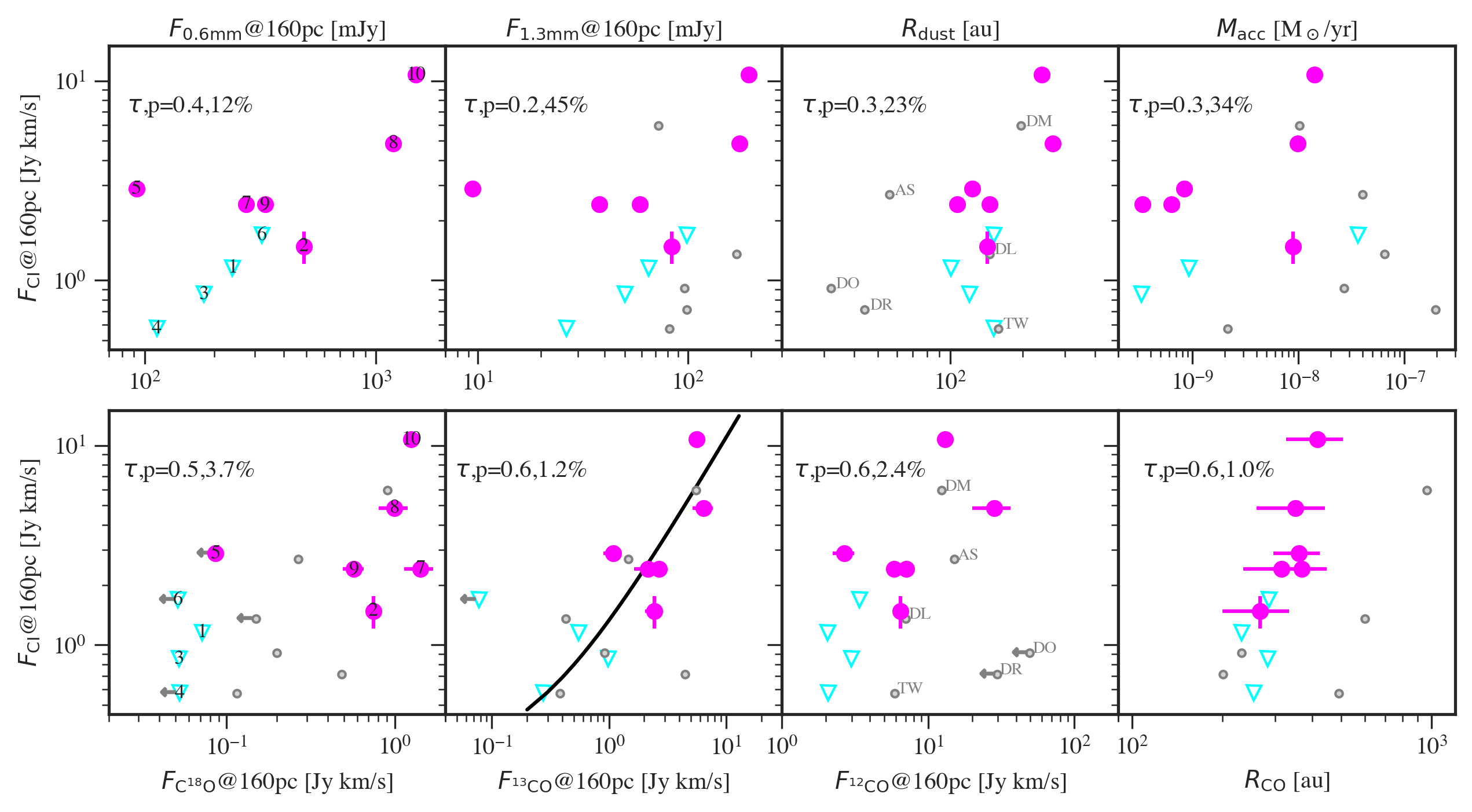}
    \caption{\CI\ line fluxes scaled at 160\,pc vs relevant disk properties. Upper panels from left to right: Band~8 continuum flux density, Band~6 continuum flux density, dust disk radius at 160\,pc, and mass accretion rate. Lower panels left to right:  C$^{18}$O\,(2-1),  $^{13}$CO\,(2-1), and $^{12}$CO\,(2-1) line fluxes at 160\,pc, and CO gas disk radius. For the Lupus sample detections are indicated with magenta filled circles while upper limits are represented by cyan triangles pointing downward. Literature T~Tauri disks with \CI\ detections are shown with gray circles. The literature $^{12}$CO fluxes from DO~Tau and DR~Tau include extended non-Keplerian disk emission, hence are treated as upper limits (see Appendix~\ref {app:lit}). The results of the \texttt{pymccorrelation} Kendall's $\tau$ tests for our Lupus sample are reported in each panel: the \CI\ flux is positively correlated with the the gas disk outer radius ($R_{\rm CO}$) as well as with the $^{13}$CO, $^{18}$O, and $^{12}$CO line fluxes. Only the first three correlations hold for the combined Lupus and literature sample, see Table~\ref{tab:stat} and Sect.~\ref{sec:CIindisks}.  The black line in the $F_{\rm CI}-F_{\rm ^{13}CO}$ panel gives the best-fit relation between these two quantities for the Lupus+literature sample, see Sect.~\ref{sec:CIindisks} for details.}
    \label{fig:CIcorr}
\end{figure*}

To further explore this inference, we search for correlations between line detections and upper limits with other star/disk properties collected in Tables~\ref{tab:sample} and~\ref{tab:res}, scaling all values to a reference distance of 160\,pc.  The upper panels of Figure~\ref{fig:CIcorr} show relations with quantities tracing the dust continuum emission ($F_{\rm 0.6mm}$ and $F_{\rm 1.3mm}$), the dust radial extent ($R_{\rm dust}$), and the stellar mass accretion rate ($\dot{M}_{\rm acc}$).  The lower panels summarize the relations with quantities probing the gas content ($F_{\rm C^{18}O}$, $F_{\rm ^{13}CO}$, and $F_{\rm ^{12}CO}$  line fluxes for the 2-1 transition) and gas radial extent ($R_{\rm CO}$). Given the significant number of \CI\ non-detections in our Lupus sample and the large errorbars for some of the quantities (e.g., $R_{\rm CO}$), we use the \texttt{pymccorrelation} routine v0.2.5\footnote{At the time of submission, there was an error in \texttt{pymccorrelation} that was patched locally. This edit can be seen at https://github.com/privong/pymccorrelation/compare/ \\ pymccorr...Bennett-Skinner:pymccorrelation:patch-1.} \citep{Privon2020ApJ...893..149P} to carry out the generalized non-parametric Kendall's $\tau$ test and investigate whether the aforementioned  stellar/disk properties are correlated with the \CI\ emission. 
The Kendall’s $\tau$ for uncensored data is calculated from two matrices, $a$ and $b$, $a_{ij}$ is $-1$ if $X_i > X_j$, $0$ (or uncertain) if $X_i = X_j$, and $1$ if $X_i < X_j$, where $X_i$ is the ith value of the independent variable; $b_{ij}$ is calculated similarly for the dependent variable. To include non-detections \texttt{pymccorrelation} adopts the method of \cite{Isobe1986ApJ...306..490I}: if $X_j$ is an upper limit, it is considered less than $X_i$ ($a_{ij} = -1$), only when $X_i > X_j$ and $X_i$ is a detection or lower limit, see \cite{Isobe1986ApJ...306..490I} for a full description of the methodology. Measurement uncertainties are accounted for with an Monte Carlo approach that randomly draws every data point independently from a Gaussian with a mean and standard deviation of its reported value and error \citep{Curran2014arXiv1411.3816C}. For each pair of variables, we ran 10,000 tests using \texttt{pymccorrelation} and report in Table~\ref{tab:stat} the
median value of Kendall's $\tau$, a value running from -1 to 1 indicating the direction of the correlation, and $p$, the percent probability that  two quantities are uncorrelated. 
The frequency distribution of Kendall's $\tau$ from these tests is not necessarily Gaussian, so the median value may differ from the value obtained when not accounting for the uncertainty in the data, hence our choice of reporting also the 16th and 84th percentile values of $\tau$.  The large uncertainties in the value of $\tau$ indicate the need for further observations, however, we stress that in every instance where the median value of $\tau$ indicates significance, barring the already marginal $F_{\rm C^{18}O}$ correlation, the 16th and 84th percentile values do as well.

\begin{deluxetable}{lcc|cc}
\tablecaption{Summary of the \texttt{pymccorrelation} Kendall's $\tau$ tests.  \label{tab:stat}}
\tablewidth{0pt}
\tablehead{
\colhead{Quantity} &  \multicolumn{4}{c}{$F_{\rm CI}$ @\,160\,pc} \\
\colhead{@\,160\,pc} & \multicolumn{2}{c|}{Lupus} & \multicolumn{2}{c}{Lupus+lit.} \\
    & \colhead{$\tau$(16th,84th)} & \colhead{$p$(16th,84th)}  & \colhead{$\tau$(16th,84th)} & \colhead{$p$(16th,84th)}
}
\startdata
$F_{\rm 0.6mm}$ & 0.39(0.37,0.42) & 12(9,13) & $-$ & $-$ \\
$F_{\rm 1.3mm}$ & 0.19(0.17,0.23) & 45(35,51) & 0.18(0.16,0.21) & 32(26,37) \\
$R_{\rm dust}$  & 0.30(0.14,0.44) & 23(8,52) & 0.25(0.23,0.28) & 17(13,22) \\
$\dot{M}_{\rm acc}$ & 0.26(0.20,0.29) & 34(28,45) & 0.03(0,0.05) & 88(80,96) \\
$F_{\rm C^{18}O}$ &  {\bf 0.52}(0.44,0.59) & {\bf 3.7}(1.8,7.6) & {\bf 0.43}(0.30,0.50) &  {\bf 2.1}(0.7,11) \\
$F_{\rm ^{13}CO}$ & {\bf 0.62}(0.57,0.67) & {\bf 1.2}(0.7,2.2) & {\bf 0.51}(0.47,0.54) &  {\bf 0.6}(0.4,1.2) \\
$F_{\rm ^{12}CO}$ & {\bf 0.56}(0.51,0.58) & {\bf 2.4}(2.0,3.9) & 0.30(0.28,0.33) &  10(7.5,13) \\
$R_{\rm CO}$  & {\bf 0.64}(0.51,0.75) & {\bf 1.0}(0.3,3.9) & {\bf 0.44}(0.41,0.47) &  {\bf 2.2}(1.5,3.2) 
\enddata
\tablecomments{Median values, 16th, and 84th percentiles for $\tau$ and $p$. $\tau$ gives the direction of the correlation (positive for $\tau > 0$) while $p$ is the percent probability that two quantities are uncorrelated. Entries with $p$ less than 5\% indicate a likely correlation, hence are in boldface.}
\end{deluxetable}

Restricting ourselves to the Lupus sample, we find that the \CI\ emission is likely positively correlated with the gas outer radius ($R_{\rm CO}$) and with the $^{13}$CO, C$^{18}$O, and $^{12}$CO\,(2-1) emission ($F_{\rm ^{13}CO}$, $F_{\rm C^{18}O}$, and $F_{\rm ^{12}CO}$), hence similarly probing the disk surface. However, only the first three correlations persist  when adding to our Lupus sample 6 more T~Tauri sources from different star-forming regions that have \CI\ detections likely tracing a disk (see Appendix~\ref{app:lit} for details on these sources, gray symbols in Figure~\ref{fig:CIcorr}, and the last columns of Table~\ref{tab:stat}). The absence of a correlation with the $^{12}$CO emission in the extended sample might be due to the main CO isotopologue being more affected by cloud absorption (\citealt{Ansdell2018} and Ansdell priv. comm.)  and sometimes tracing extended structures unrelated to the circumstellar disk, e.g. envelopes and outflows \citep[e.g.,][]{Kurtovic2018ApJ...869L..44K,Huang2023ApJ...943..107H}.  
C$^{18}$O exhibits a weaker correlation with \ci\ than $^{13}$CO, likely  because it probes gas closer to the disk midplane \citep[e.g.,][and Sect.~\ref{sec:modres}]{Miotello2016A&A...594A..85M,Law2021ApJS..257....4L,Ruaud2022ApJ...925...49R,Kama2016theory}. We also note that the $F_{\rm CI}$ and $F_{\rm ^{13}CO}$ follow very closely a one-to-one linear relation. Indeed, when 
using \texttt{linmix} \citep{Kelly2007ApJ...665.1489K} to account for upper limits and uncertainties on the
Lupus+literature sample\footnote{We exclude Sz~98 (ID~6) since it is not detected in either of the lines.} we find  $F_{\rm CI}= 1.07(\pm 0.33) \times \, F_{\rm ^{13}CO} + 0.26(\pm 0.89)$  where fluxes are in Jy\,km/s (black line in the $F_{\rm CI}-F_{\rm ^{13}CO}$ panel of Figure~\ref{fig:CIcorr}).
Finally, the lack of correlations with  dust properties, in the Lupus sample as well as in the extended sample, demonstrates that the \CI\ emission is not affected by the amount or radial extent of mm-sized grains which mostly trace icy pebbles in the disk midplane \citep[e.g.,][]{Villenave2020A&A...642A.164V}.  In conclusion, empirical evidence from the \CI\  profiles and  correlations with other disk tracers strongly suggest that the \CI\ emission probes gas at the disk surface. We will further test this inference in the next sub-sections by comparing our observations more directly to theoretical predictions.

\subsection{Comparison with the RGH22 thermochemical disk models} \label{sec:modres}
Recently, RGH22 carried out a grid of thermochemical disk models adopting ISM carbon and oxygen elemental abundances as input parameters. In addition to isotopologue-selective photodissociation, they added three-phase grain-surface chemistry with CO conversion into CO$_2$ ice being a major reaction and adopted vertical hydrostatic equilibrium
 to derive a self-consistent gas density and temperature. The CO conversion into CO$_2$ ice shifts the CO snowline vertically away from the midplane, thus reducing the amount of CO on the disk surface. 
Within a factor of a few, their predicted C$^{18}$O(3-2) luminosities match observations of Lupus and Chamaeleon~I disks detected in this line. This result led RGH22 to argue that C$^{18}$O is a good tracer of the gas disk mass and that no severe elemental or CO depletion by other chemical or dynamical processes is necessary to reconcile theoretical predictions with observations.
Here, we take the comparison a step further and test whether these same models can explain the emission from three CO isotopologues as well as the \CI\ line which is the focus of this study.

\begin{figure}
    \centering
    \includegraphics[width = 0.5\textwidth]{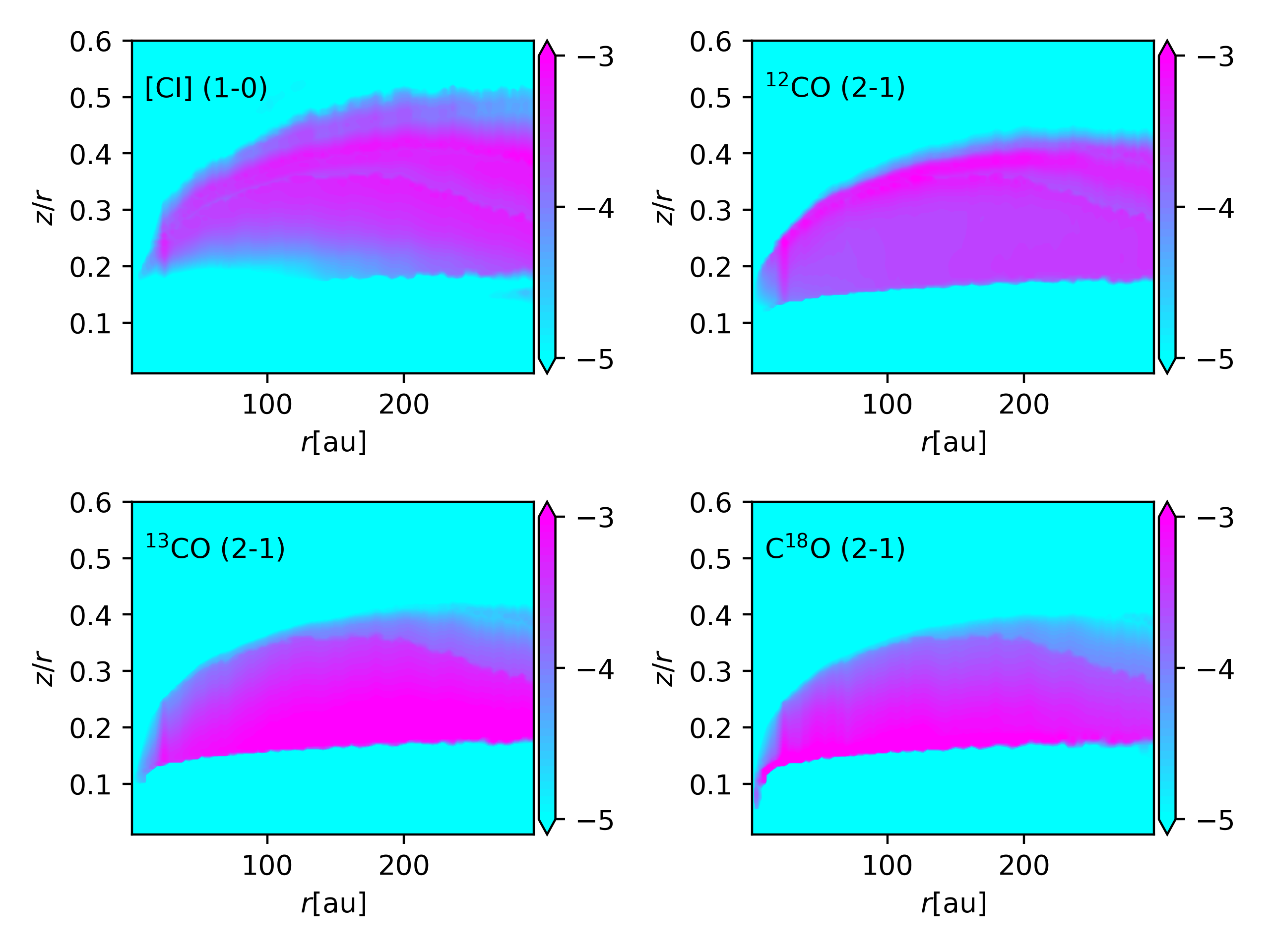}
    \caption{Normalized fractional luminosity (in log units) for the \CI\ (upper left), $^{12}$CO\, 2-1 (upper right), $^{13}$CO\, 2-1 (lower left), and C$^{18}$O 2-1 (lower right) lines. Models are from RGH22 for a disk radius of 300\,au, gas mass 0.01\,M$_{\odot}$, and gas-to-dust ratio of 100. Full non-LTE radiative transfer is carried out to compute line luminosities. \CI\ traces the disk atmosphere down to z/r$\sim 0.2$. }
    \label{fig:carbon}
\end{figure}

First, we use the RGH22 model with a disk outer radius of 300\,au, a minimum mass solar nebula (MMSN) gas  of 0.01\,M$_\odot$, and $\Delta_{\rm gd}$ of 100 (see their Table~1) to compare the emitting surfaces of various carbon species.  Figure~\ref{fig:carbon}  shows that the \CI\  line probes the uppermost surface of the disk down to $z/r\sim 0.2$, thus overlapping with the $^{12}$CO and $^{13}$CO\,(2-1) emitting surfaces, while the  C$^{18}$O\,(2-1) emission is concentrated at lower altitudes ($z/r\sim 0.1$). We  note that the predicted CO emitting surfaces agree with those empirically derived from the  ALMA MAPS survey: in five disks observed at high sensitivity and spatial resolution, $^{12}$CO\,(2-1) emission is found to be mostly at $z/r >0.3$ while  $^{13}$CO and C$^{18}$O\,(2-1) lie below, at  $z/r \approx 0.1-0.2$ \citep{Law2021ApJS..257....4L}. In the context of this study, it is worth mentioning that the column density of carbon is set by photoionization of C into C$^+$ and photodissociation of CO to C and, in agreement with \citet{Kama2016}, the \CI\ line is found to be mostly optically thin. The correlation among fluxes reported in Sect.~\ref{sec:CIindisks} could be attributed to the overlapping emitting surfaces between the \CI\ line and the $^{12}$CO, $^{13}$CO, and, to a lesser extent, C$^{18}$O (2-1) lines.

\begin{figure*}[h]
    \centering
    \includegraphics[width =\textwidth]{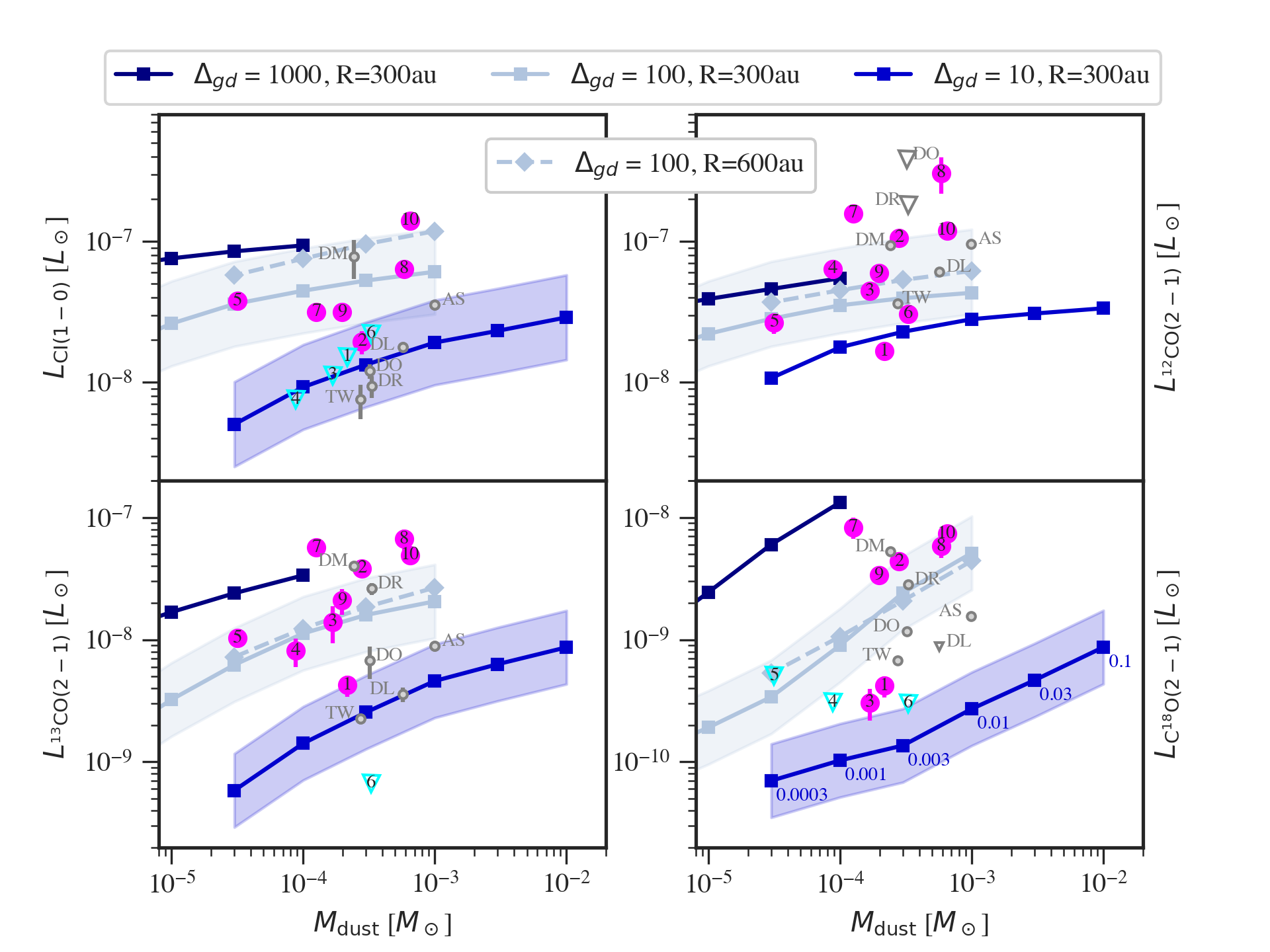}
    \caption{Comparison between the RGH22 models for an outer disk radius of 300\,au (squares) and observations (symbols as in Figure~\ref{fig:CIcorr}). The shaded regions depict a range of values that are within a factor of two from model predictions.  For each gas-to-dust ratio ($\Delta_{gd}$), the sequence of models indicates a different total disk mass (see labels in the bottom right panel next to the squares of the $\Delta_{gd} = 10$ track where the total disk mass is in solar masses).
    The diamond track is for the same surface density disk with $\Delta_{gd} = 100$ but the outer radius is extended to 600\,au, illustrating that the  \CI\ and $^{12}$CO\,(2-1) lines are sensitive to the disk outer radius.  All Lupus disks have CO radii $\sim 200-400$\,au, with V1094~Sco and IM~Lup (ID~8 and 10) being the largest in this sample. Among the literature sources (gray symbols) the largest disk is that of DM~Tau ($\sim 850$\,au)  while the smallest disks are those of TW~Hya and DR~Tau ($\sim 180$\,au), see Table~\ref{tab:others}. }
    \label{fig:compare}
\end{figure*}

Next, we carry out a direct comparison of predicted and observed  \CI\ and CO luminosities vs. dust disk masses ($M_{\rm dust}$), see Figure~\ref{fig:compare}. The grid models for an outer disk radius of 300\,au (squares) are the same as presented in RGH22 and cover a large range in disk mass (from $3 \times 10^{-4}$ to 0.1\,M$_\odot$) and three gas-to-dust mass ratios ($\Delta_{\rm gd} = $10, 100, 1000). 
To test whether the adopted outer radial cutoff captures most of the emission, we also run 4 models for $\Delta_{\rm gd} = 100$ where we extend the radial grid to 600\,au (grey diamonds connected by dashed lines in Figure~\ref{fig:compare}). This test demonstrates that the  $^{13}$CO and  C$^{18}$O\,(2-1) emission is confined within 300\,au while only $\sim$\,50\% of \CI\ emission is contained within this radius. Note that in all models the primary input parameters are the dust surface density, the dust mass, and the gas-to-dust mass ratio while the gas structure is computed by solving for vertical hydrostatic pressure equilibrium. However, we compare predicted line luminosities vs. dust disk masses because the latter are constrained by observations. In carrying out this comparison, we also took into account that all models assume 
a face-on disk inclination and thus maximum emission for optically thick lines. Since the $^{12}$CO and $^{13}$CO lines are expected to be optically thick we divide the observed fluxes by the cosine of the measured disk inclination before converting them into luminosities. The dust disk mass for the Lupus+literature sample is  calculated from 1.3\,mm fluxes (Tables~\ref{tab:sample} and \ref{tab:others}) assuming optically thin emission \citep[e.g., eq.~2 in][]{Pascucci2016ApJ...831..125P}, a dust temperature of 20\,K, and a dust opacity at 1.3\,mm of 1.5\,cm$^2$/g, instead of the 2.3\,cm$^2$/g typically adopted in observational papers \citep[e.g.,][]{Andrews2013}, to  match the RGH22 dust properties. In summary, the panels shown in  Figure~\ref{fig:compare} constrain the dust mass (through the mm flux density), gas mass (through the  C$^{18}$O line when detected or the $^{13}$CO line otherwise), and carbon content (through the \CI\ flux) of a disk. Furthermore, the $^{12}$CO\,(2-1) and \CI\ fluxes are  sensitive to the gas outer radius.

We start by commenting on the  disk mass and $\Delta_{\rm gd}$. The Lupus+literature sample covers more than an order of magnitude in $M_{\rm dust}$. All disks with a C$^{18}$O(2-1) detection lie above the RGH22 $\Delta_{\rm gd}=10$ track, half of them are actually above $\Delta_{\rm gd}=100$  (bottom right panel in Figure~\ref{fig:compare}). For sources with a C$^{18}$O upper limit, perhaps indicative of a low gas mass, we can use the $^{13}$CO luminosities (bottom left panel in Figure~\ref{fig:compare}) to gauge their gas content. 
With the exception of Sz~98 (ID~6), all the sources are at or above the $\Delta_{\rm gd}=10$ track, with the Lupus disks being closer to or above $\Delta_{\rm gd}=100$. 
The $^{13}$CO upper limit from Sz~98 is a factor of a few below the $\Delta_{\rm gd}=10$ track, hence this disk might have experienced significant (more than a factor of 10) gas or CO depletion. Deeper observations of the rare CO isotopologues as well as other gas mass tracers (e.g., N$_2$H$^+$ \citealt{Anderson2019ApJ...881..127A}) would be useful to pin down the extent and origin of this depletion. Among the literature sources, DL~Tau is the only one with an C$^{18}$O upper limit and its $^{13}$CO\,(2-1) and \CI\ fluxes point to a depletion in gas mass (or carbon) of a factor of 10. Still, its gas mass is $> 0.003$\,M$_\odot$ which is about three times the mass of Jupiter.  In summary, based on the data at hand and the RGH22 models, all of the disks investigated here, except  Sz~98, have more than enough mass to form a Jupiter mass planet. Some of them, like J16083070 (ID~7), V1094~Sco (ID~8), and IM~Lup (ID~10), have disks as massive as $\sim$0.1\,M$_\odot$, i.e. ten times the MMSN. This result agrees with and expands upon  what was inferred in RGH22 where the comparison was restricted to the Lupus and Chamaeleon~I disks with C$^{18}$O\,(3-2) detections.

The upper two panels of Figure~\ref{fig:compare} cover lines that trace the uppermost disk surface and are most sensitive to the gas disk outer radius. J16083070 (ID~7), V1094~Sco (ID~8), IM~Lup (ID~10), and DM~Tau are the largest disks and, indeed, among the strongest emitters in the $^{12}$CO\,(2-1) and \CI\ lines. On the opposite end, TW~Hya and DR~Tau have the smallest CO gas disk radii ($\sim 180$\,au) and the lowest \CI\ luminosities, a factor of $\sim 5$ below the $\Delta_{\rm gd}=100$ track for a gas disk radius of 300\,au. Even considering their smaller gas disk radii, a depletion in carbon of a factor of a few may be needed to explain their low \CI\ luminosities. A similar conclusion has been reached in RGH22 for TW~Hya with a disk model tailored to this source that can also reproduce the $^{13}$CO\,(2-1) flux with $\Delta_{\rm gd}=100$, see their Fig.~8. This highlights the importance of target-specific modeling, see also Deng et al.~2023 in press (arXiv:4990967). The \CI\ luminosities from RY~Lup (ID~2), J16000236 (ID~3), Sz~133 (ID~4), and DO~Tau also indicate a factor of a few to several depletion in carbon: for ID~3 and 4 there could also be an overall factor of a few depletion in CO or gas based on their C$^{18}$O fluxes (see Figure~\ref{fig:compare} bottom right panel). In contrast, for Sz~91 (ID~5), J16083070 (ID~7), V1094Sco (ID~8), Sz~111 (ID~9), IM~Lup (ID~10), and DM~Tau the lines investigated here do not indicate any depletion in gas, CO, or carbon and, within a factor of a few, are consistent with the  $\Delta_{\rm gd} \ge 100$ tracks.

At this point it is useful to comment on which star and disk parameters might affect most the \CI\ line, hence our inference of negligible carbon depletion. As mentioned in Sect.~\ref{sec:intro}, atomic carbon forms above the CO photodissociation layer and below the C ionization front. In that layer, UV attenuation is determined by a combination of carbon absorption and by dust. Indeed, we can see in Figure~\ref{fig:compare} that, for a fixed gas mass, changing the gas/dust ratio by a factor of 100 changes the \CI\ luminosity by a factor of $\sim 10$. This means that the amount of dust, along with its degree of settling, affects the abundance of carbon at the disk surface. On the opposite, there is only a modest dependence with gas mass: Following one of the $\Delta_{\rm gd}$ tracks in Figure~\ref{fig:compare}, one sees that  changing the gas mass by a factor of 100 changes \CI\ luminosity only by a factor of a few. Results are also not sensitive to different cosmic ray ionizations \citep[e.g.,][]{Kama2016} as the cosmic ray ionization rate is much lower than UV photorates at the surface. In addition, the \ci\ emission is also not sensitive to the overall UV flux because it always arises from an approximately fixed column corresponding to a few UV optical depth  \citep[e.g.,][]{Kaufman1999ApJ...527..795K}. This is why the \CI\ line has been chosen in this and previous studies as a suitable probe for carbon depletion. 

\subsection{Comparison with results from the literature}\label{sec:complit}
Of the 16 Lupus+literature disks discussed in this paper, 11 have previously reported gas and dust disk masses, hence $\Delta_{\rm gd}$, while for 5 there are  literature constraints on their C/H elemental abundance ratio. 

We start by discussing the first group of 11 disks where gas mass estimates have been obtained by matching observed to predicted CO isotopologue fluxes: a) for 7 sources using a grid  of physical-chemical disk models obtained with DALI \citep{Bruderer2012A&A...541A..91B}, see \citet{Miotello2017A&A...599A.113M}; b) for Sz~71 \citep[ID~1][]{Ansdell2018} and DM~Tau using the grid of parametric disk models by \citet{WilliamsBest2014}; c) for TW~Hya and IM~Lup (ID~10) by generating individual disk models \citep{Favre2013ApJ...776L..38F,Zhang2021ApJS..257....5Z}. It is worth mentioning that among these approaches only a) includes isotope-selection dissociation which, according to \citet{Miotello2014A&A...572A..96M}, can decrease the optically thin emission of C$^{18}$O by an order of magnitude. In addition, approach a), b), and the individual modeling of TW~Hya by \cite{Favre2013ApJ...776L..38F} do not include CO conversion to CO$_2$ ice which, according to \citet{Trapman2021A&A...649A..95T} and \citet{Ruaud2022ApJ...925...49R}, can  further decrease the  C$^{18}$O flux by a factor of a few. Therefore, it is not surprising that the literature C$^{18}$O  model fluxes are larger than observed and significant gas or CO depletion had to be invoked to reconcile models with observations. For instance, ID~2, 3, 5, 7, and 9 have literature $\Delta_{\rm gd} \sim 3-10$ \citep{Miotello2017A&A...599A.113M} while according to the RGH22 grid only ID~3 lies clearly below the  $\Delta_{\rm gd} = 100$ track and only by a factor of a few. Even lower  $\Delta_{\rm gd}$ ($\leq 1$) have been reported for ID~1, 4, 6, 10, and TW~Hya \citep{Miotello2017A&A...599A.113M,McClure2016ApJ...831..167M,Zhang2021ApJS..257....5Z}. Among this group, only ID~6 (Sz~98) could be depleted according to RGH22 but, given the current $^{13}$CO upper limit, only by a factor slightly larger than $\sim 10$,  significantly less than what reported in the literature.
The most discrepant result is that for IM~Lup (ID~10), a highly accreting star surrounded by the largest gaseous disk in Lupus. \cite{Zhang2021ApJS..257....5Z} 
used RADMC3D \citep{Dullemond2012ascl.soft02015D} to fit the spectral energy distribution of IM~Lup and constrain the disk structure, including the gas and dust density and dust temperature profiles. Next, they ran the chemical code RAC2D \citep{Du2014ApJ...792....2D} for 1\,Myr to obtain the gas temperature and chemical abundances and finally ran RADMC3D again to obtain $^{13}$CO and C$^{18}$O (2-1) and (1-0) cubes to be compared with the MAPS ALMA datacubes \citep{Oberg2021ApJS..257....1O}. 
\cite{Zhang2021ApJS..257....5Z} can only reproduce the CO column density radial profile for IM Lup when reducing the CO gas abundance by two orders of magnitude with respect to the ISM value of $\sim 10^{-4}$. However, as mentioned in \cite{Zhang2021ApJS..257....5Z}, such a large CO depletion cannot be reached for this young ($\sim 1$\,Myr, \citealt{Alcala2017}) disk even when combining disk chemical processes with turbulent mixing and sequestration of CO ice in the disk midplane \citep{Krijt2020ApJ...899..134K}. We want to emphasize that, based on the RGH22 grid, a significant depletion of CO is not required to explain the  integrated $^{13}$CO and  C$^{18}$O fluxes of IM~Lup. Rather, the physical and chemical processes that are included in this grid of models (e.g., freeze-out, selective dissociation, CO conversion into CO$_2$ ice, and vertical hydrostatic equilibrium) are sufficient to reproduce the CO isotopologue fluxes as well as the high  \CI\ flux (Figure~\ref{fig:compare}). According to these models, the disk of IM~Lup can have an ISM gas-to-dust ratio of 100 and is more massive than the MMSN. Interestingly, a similarly high gas disk mass can be independently estimated from the right panel of Figure~7 in \citet{Miotello2016A&A...594A..85M} without invoking any extra CO depletion beyond freeze-out and selective photodissociation. It is worth  re-stating that RAC2D does not include isotope-selective photodissociation. In addition, it was used in \cite{Zhang2021ApJS..257....5Z} mostly to obtain a stable temperature profile and, when varying the CO gas abundance, the chemistry was not rerun. On the other hand, the comparison here is restricted to integrated line fluxes. Dedicated self-consistent gas and dust models of IM~Lup would be extremely valuable to evaluate the extent of any radial CO depletion.

Fewer T~Tauri stars have been observed in the \CI\ line than in the main CO isotopologues and, before this study, only 6 sources had a reported detection likely arising from the disk, see Table~\ref{tab:others} in Appendix~\ref{app:lit}. Among these literature sources, the \CI\ and CO isotopologue emission from DL~Tau, DM~Tau, DO~Tau, DR~Tau, and TW~Hya were  modeled using the DALI code \citep{Kama2016,Sturm2022}. These works report carbon depletion factors with respect to ISM values of $\sim 160$, 5, 15, 5, and 100, respectively. 
Caution should be taken for the Taurus sources as flux loss of a factor of several and up to an order of magnitude affects the $^{13}$CO and C$^{18}$O data used in \citet{Sturm2022}, Sturm priv. comm. This is why here we adopted literature values (see Table~\ref{tab:others}). Based on these values, we find that the generic RGH22 models do not require orders of magnitude depletion in carbon. Even for DL~Tau and TW~Hya the RGH22 grid suggests carbon depletion much lower than 100, with factors of just ten and a few, respectively. These more modest depletions can be easily accounted for through chemical \citep[e.g.,][]{Schwarz2018ApJ...856...85S} and/or dynamical processes \citep[e.g.,][]{Krijt2018ApJ...864...78K}.

\section{Summary and Outlook} \label{sec:summary}
We have acquired and analyzed ALMA/ACA Band~8 data covering the \CI\ line at 492.161\,GHz for 9 large gaseous disks ($R_{\rm CO} \gtrsim 200$\,au) around T~Tauri stars in the $\sim 1-3$\,Myr-old Lupus star-forming region. We have also retrieved and analyzed archival ALMA/12-m Band~8 data for IM~Lup whose disk has a CO radius of $\sim 400$\,au, the largest in the region. Our Lupus sample covers a factor  of $\sim 20$ in 1.3\,mm flux density, hence likely dust disk mass. Finally, to place our Lupus sample into context, we have assembled literature source properties for T~Tauri stars with a \CI\ detection likely arising from a disk, an additional 6 sources. Our results can be summarized as follows:
\begin{itemize}
    \item Band~8 continuum emission is detected towards all Lupus disks and it is confined within the large ACA beam ($3.2 '' \times 2 ''$) for all sources except for V1094~Sco which is marginally resolved. The continuum emission from IM~Lup is clearly resolved with the smaller beam ($0.37 '' \times 0.32 ''$) of the archival 12-m data.  These results are in line with already published  1\,mm continuum observations and analysis.
    \item The \CI\ line is detected in 6 out of 10 Lupus sources with centroids and FWHMs consistent with outer gas ($\gtrsim 100$\,au) in a Keplerian disk: the profiles from IM~Lup and J16083070 are clearly double peaked. Thus, our work doubles the sample of \CI\ detections from T~Tauri disks. All six \CI\  detections are from large CO disks, $R_{\rm CO} \gtrsim 250$\,au.
    \item The \CI\ emission is not correlated with the dust emission ($F_{\rm 0.6 mm}$ and $F_{\rm 1.3 mm}$) or its radial extent ($R_{\rm dust}$). Instead, it is correlated with the gas radial extent ($R_{\rm CO}$), the $^{12}$CO, C$^{18}$O, and $^{13}$CO emission, most tightly with the $^{13}$CO\,(2-1) flux. The correlations with $R_{\rm CO}$ and the rare CO isotopologue fluxes persist when adding to the Lupus sample the six additional T~Tauri stars with \CI\ detections from the literature.
\end{itemize}

When comparing the inferred \CI\ and the $^{12}$CO, $^{13}$CO, and C$^{18}$O (2-1) luminosities to those predicted by RGH22, we find no evidence for significant gas, CO, or carbon depletion in our Lupus sample except for Sz~98. This disk may be depleted in gas or CO by a factor $\gtrsim 10$, deeper  observations are needed to place  firm constraints. Importantly, the integrated line luminosities from IM~Lup, a highly accreting star with the largest gaseous disk in the region, are fully consistent with a massive gaseous disk ($\sim 0.1$\,M$_\odot$) without any CO or carbon depletion beyond what is set by freeze-out,  CO conversion into CO$_2$ ice, and isotope-selective photodissociation. 
Our conclusion applies to all literature sources with \CI\ detections, including TW~Hya, with the exception of DL~Tau.
For DL~Tau, it may be necessary to consider a depletion (or gas or CO or carbon)  of up to a factor of 10.

In contrast to the conclusions driven above, several past works have claimed large carbon and/or CO depletion in disks around T~Tauri stars \citep[e.g.,][]{WilliamsBest2014,Kama2016theory,Miotello2016A&A...594A..85M}. Specifically for IM~Lup, a reduction in CO of a factor of 100  has been reported to explain its column density radial profile  \citep{Zhang2021ApJS..257....5Z}.
Some of these inconsistencies appear to arise from inadequate millimeter observations, which are either too shallow or lack the necessary short baselines to detect the entire flux emitted by these large disks. This issue is exemplified by the case of V1094~Sco (Sect.~\ref{sect:obs_and_redu}) and the Taurus literature sources discussed in this paper (Sect.~\ref{sec:complit}). Additionally, we have speculated that some of the discrepancies may stem from missing physics in the  chemical models used to interpret the data (e.g., isotope-selective dissociation and CO conversion to CO$_2$ ice, see also \citealt{Ruaud2022ApJ...925...49R} and \citealt{Trapman2021A&A...649A..95T}), as well as a lack of self-consistent dust and gas modeling.  
Efficient conversion of CO into CO$_2$ ice could be investigated via JWST/NIRspec and MIRI-MRS spectroscopy of selected edge-on disks. Along with retrieving the relative column densities of CO and CO$_2$ ice,  the shape of the CO$_2$ absorption features at $\sim 4.2$ and 15\,\micron{}  \citep[e.g.,][]{McClure2023NatAs...7..431M} may indicate formation on a water-ice coated grain, as predicted by the RGH22 models.
Detailed Benchmark tests should  be also carried out to resolve any large discrepancies between model predictions. Additionally, dedicated self-consistent gas and dust models should be developed for disks with spatially resolved CO isotopologue profiles in order to evaluate the degree of any radial CO depletion. 
Meanwhile, our analysis, which relies on integrated line fluxes, 
indicates that large Myr-old disks may conform to the straightforward expectation that they are not substantially depleted in gas, CO, or carbon.

\begin{acknowledgments}
The authors thank N. Kurtovic, F. Long, J. A. Sturm, and S. van Terwisga for information shared on specific sources which were not readily available from published papers. The authors also thank the AGE-PRO calibration team for sharing the CO isotopologue fluxes of Sz~71 and V1094~Sco in advance of publication. I.P. thanks the NAASC Staff, in particular Sarah Wood, for help with the initial ACA data reduction. I.P., D.D., and U.G. acknowledge support from the NASA/XRP research grant 80NSSC20K0273. Support for M.R.’s research was provided by NASA’s Planetary Science Division Research Program, through ISFM work package ‘The Production of Astrobiologically Important Organics during Early Planetary System Formation and Evolution’ at NASA Ames Research Center.
This paper makes use of the following ALMA data: 2019.1.00927.S and 2015.1.01137.S. ALMA is a partnership of ESO (representing its member states), NSF (USA), and NINS (Japan), together with NRC (Canada), MOST and ASIAA (Taiwan), and KASI (Republic of Korea), in cooperation with the Republic of Chile. The Joint ALMA Observatory is operated by ESO, AUI/NRAO, and NAOJ.
\end{acknowledgments}

\facilities{ALMA(ACA)}

\software{AstroPy \citep{Astropy2013A&A...558A..33A}, CASA \citep{McMullin2007ASPC..376..127M}, matplotlib \citep{Hunter2007CSE.....9...90H}, Scipy (http://www.scipy.org), pymccorrelation \citep{Privon2020ApJ...893..149P}, specutils \citep{Earl2022zndo...7348235E}}

\appendix
\section{ALMA Observing Log}\label{app:ALMAlog}
Our Band~8 ACA proposal (2019.1.00927.S, PI: I. Pascucci) was accepted in July 2019 with a priority grade of B. Unfortunately, the COVID-19 pandemic and subsequent shutdown of the ALMA facility prevented achieving the requested sensitivity. Despite these challenges, we are grateful for the dedicated efforts of the ALMA observatory, which enabled acquiring valuable data over the span of $\sim 1.5$ years. Table~\ref{tab:ALMAlog} summarizes the number of antennas and  integration time per observing block. Although all of the targets were observed in each observing block, the total on-source integration times are not identical and vary from 37.30\,min for Sz~91 to 56.11\,min for V1094~Sco. The other on-source integration times are as follows: 50.74\,min for Sz~71; 48.22\,min for RY~Lup; 51.41\,min for J16000236; 45.70\,min for Sz~133; 47.71\,min for Sz~98; 45.70\,min for J16083070; and 42.17\,min for Sz111. 

\begin{deluxetable}{cccccccc}
\tablecaption{ALMA Observing Log\label{tab:ALMAlog}}
\tablewidth{0pt}
\tablehead{
\colhead{Execution Blocks} & \colhead{N$_{\rm ant}$} & \colhead{Calibrators} &  \colhead{Integration Time} \\
\colhead{(UTC Time)} & \colhead{} & \colhead{} &  \colhead{(s)}  
}
\startdata
2020-2-28 9:42:07 & 10 & J1604-4441, J1610-3958, J1924-2914 &  42:41 \\
2021-6-13 3:47:11 & 8 & J1514-4748, J1604-4441, J1626-2951, J1924-2914 &  10:05 \\
2021-7-01 1:21:08 & 8 & J1514-4748, J1517-2422, J1604-4441 &  43:21 \\
2021-7-04 23:59:22 & 9 & J1337-1257, J1514-4748, J1604-4441 &  42:21 \\
2021-7-05 2:15:02 & 9 & J1604-4441, J1650-5044, J1924-2914 &  42:21 \\
2021-7-05 23:14:04 & 9 & J1337-1257, J1514-4748, J1604-4441 &  42:21 \\
2021-7-08 2:30:07 & 8 & J1604-4441, J1650-5044, J1924-2914 &  43:21 \\
2021-7-09 00:48:46 & 9 & J1514-4748, J1517-2422, J1604-4441 &  43:21 \\
2021-7-09 2:57:34 & 9 & J1604-4441, J1650-5044, J1924-2914 &  42:21 \\
2021-7-10 23:27:37 & 10 & J1337-1257, J1514-4748, J1604-4441 &  42:21 \\
2021-7-11 1:43:52 & 10 & J1604-4441, J1610-3958, J1924-2914 &  42:21 \\
2021-8-10 1:01:01 & 8 & J1514-4748, J1604-4441, J1924-2914 &  43:51 \\
2021-8-21 00:51:23 & 8 & J1514-4748, J1604-4441, J1924-2914 & 43:21 
\enddata
\tablecomments{All targets are observed in each execution block but on-source integration times are different, see main text}
\end{deluxetable}

\section{Continuum Emission for the ACA Lupus Sample}\label{app:Continuum}
\begin{figure}
    \centering
    \includegraphics[width = 0.5\textwidth]{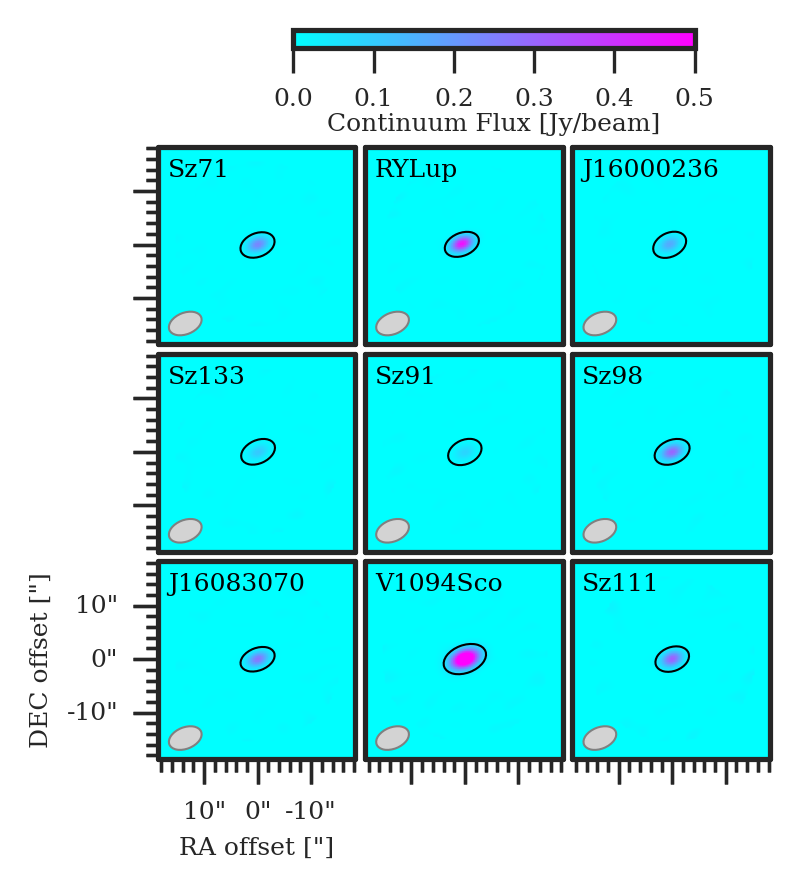}
    \caption{ALMA Band~8 self-calibrated continuum images with best fit 2D Gaussian (black ellipse) from \texttt{imfit}. The ACA beam is shown with a grey ellipse in the bottom left of each panel. The emission from V1094~Sco is marginally extended.}
    \label{fig:continuum}
\end{figure}

A gallery of the self-calibrated Band~8 continuum images for our ACA Lupus sample is provided in Figure~\ref{fig:continuum}. The best-fit 2D Gaussian is also shown as a black ellipse in each panel. Among this sample the emission from V1094~Sco is the brightest and most spatially extended.

\section{Additional T~Tauri disks with \CI\ detections}\label{app:lit}
We have searched the literature for additional T~Tauri stars with \CI\ detections. We excluded Herbig Ae/Be stars because they have a much larger FUV luminosity than T~Tauri stars and FUV photons drive the dissociation of CO into atomic and ionized carbon, hence the detectable level of \ci\ emission. We  also excluded FM~Cha, WW~Cha, and FZ~Tau because their \CI\ profiles are not dominated by disk emission but rather by the cloud or an outflow, see \citet{Sturm2022} for details. Our search led to 6 additional T~Tauri stars with \CI\ detections likely arising from the disk \citep{Tsukagoshi2015ApJ...802L...7T,Kama2016,Sturm2022}. Object properties used in this study are summarized in Table~\ref{tab:others}. In the following, we provide a few more details about the collected data.

In relation to dust and gas disk radii, we have preferred those containing 90\% of the continuum and of the $^{12}$CO\,(2-1) flux for consistency with the Lupus sample (see Table~\ref{tab:sample}).  However, a few systems do not have such estimates. The $^{12}$CO emission from AS205~N is very complex and, by extending to the southern component, likely traces tidally-stripped gas (see Fig.~5 in \citealt{Kurtovic2018ApJ...869L..44K}), hence a gas disk radius cannot be  determined. The only observations available in the ALMA archive for $^{13}$CO and C$^{18}$O are from \citet{Salyk2014} but are too low angular resolution to obtain a proper estimate. For DO~Tau the only $R_{\rm dust}$ available in the literature is the one encompassing 68\% of the millimeter flux \citep{Tripathi2017} while $R_{\rm CO}$ is the radius at half-maximum intensity \citep{KoernerSargent1995}. Finally, the only gas disk radius available for DR~Tau is the one inferred from modeling  the $^{13}$CO and C$^{18}$O emission \citep{Braun2021}, hence likely represents a lower limit for $R_{\rm CO}$. 

Although CO isotopologue fluxes for DL~Tau, DO~Tau, and DR~Tau are also available from \citet{Sturm2022}, there are concerns that these measurements may underestimate the total flux by a significant factor (Strum priv. comm.).  In light of this, we have opted to utilize literature fluxes from observations that incorporate short baselines for our study.



\begin{deluxetable*}{llccccccccccccc}
\tablecaption{Literature T~Tauri stars with \CI\ disk emission. \label{tab:others}}
\tabletypesize{\scriptsize}
\tablewidth{0pt}
\tablehead{
\colhead{ID} & \colhead{Source} & \colhead{Region} & \colhead{Dist} & \colhead{$M_*$} & \colhead{Log$\dot{M}_{\rm acc}$} & \colhead{$F_{\rm [CI]}$} & \colhead{F$_{\rm 1.3mm}$} & \colhead{R$_{\rm dust}$} & \colhead{i} & \colhead{$F_{\rm C^{18}O}$} & \colhead{$F_{\rm ^{13}CO}$} & \colhead{$F_{\rm ^{12}CO}$} & \colhead{$R_{\rm CO}$} & \colhead{Ref} \\
\colhead{} & \colhead{} & \colhead{} & \colhead{(pc)} & \colhead{(M$_\odot$)} & \colhead{(M$_\odot$/yr)} &\colhead{(Jy\,km/s)} & \colhead{(mJy)} & \colhead{($''$)} &  \colhead{(deg)} & \colhead{(Jy\,km/s)} & \colhead{(Jy\,km/s)} & \colhead{(Jy\,km/s)} &  \colhead{($''$)} & \colhead{}
}
\startdata
AS & AS205~N & Ophiuchus & 142 & 0.9& -7.4 &3.42& 377 & 0.35 & 15 & 0.34 & 1.86 & 19.23 & -- & 1,2,3,4 \\
DL & DL~Tau & Taurus& 159.94 & 0.7& -7.2 &1.36 & 170.72& 0.91 & 45 &  $<$0.15 & 0.43 & 7.05 & 3.75  &1,2,5,6 \\
DM & DM~Tau & Taurus & 144.05 & 0.3& -8.0 &7.35& 89.4& 1.23 & 36 & 1.12 & 6.84 & 15.21 & 6.04 &1,7,5,8,9 \\
DO & DO~Tau & Taurus& 141 & 0.5 & -7.6 &1.18& 123.76& 0.20\tablenotemark{\scriptsize{a}} & 37 & 0.26 & 1.18 & 63.7\tablenotemark{\scriptsize{a}} & 1.45\tablenotemark{\scriptsize{a}} &1,2,10,6,8,9,11 \\
DR & DR~Tau & Taurus & 141 & 0.6 & -6.7 &0.92& 127.18& 0.28 & 5.4 & 0.62 & 5.73  & 37.9\tablenotemark{\scriptsize{b}} & 1.26\tablenotemark{\scriptsize{b}} &1,2,11,12,13 \\
TW & TW~Hya & TW Hydra & 60 & 0.6 & -8.7 &4.08& 580& 0.99 & 5 & 0.82 & 2.72 & 41.8 & 3.07 & 14,15,16,5 \\
\enddata
\tablecomments{$F_{\rm [CI]}$ is the flux for the \CI\ line while $F_{\rm C^{18}O}$, $F_{\rm ^{13}CO}$, and $F_{\rm ^{12}CO}$ are for the (2-1) transition. Unless noted below, $R_{\rm dust}$ is the dust disk radius encompassing 90\% of the 1.3\,mm flux density while  $R_{\rm CO}$ is the gas disk radius enclosing 90\% of the ${\rm ^{12}CO}$ (2-1) flux. The $^{12}$CO emission from AS205~N is complex, hence a gas disk  radius cannot be estimated, see Appendix~\ref{app:lit} for more info.}
\tablenotetext{a}{For DO~Tau  $R_{\rm dust}$ encompasses 68\% of the mm flux density while $R_{\rm gas}$ is from modeling the ${\rm C^{18}O}$ and ${\rm ^{13}CO}$ emission. The quoted $F_{\rm ^{12}CO}$ flux is from SMA data with a beam size 1.2$'' \times$0.9$''$ \cite{WilliamsBest2014}: It is treated here as an upper limit because it likely includes outflow emission \citep{FL2020AJ....159..171F}.}
\tablenotetext{b}{The $F_{\rm ^{12}CO}$ flux for DR~Tau includes larger scale non-Keplerian emission \citep{Huang2023ApJ...943..107H}, hence it is treated as an upper limit to the disk emission in our analysis.  $R_{\rm CO}$ is from modeling the ${\rm C^{18}O}$ and ${\rm ^{13}CO}$ emission.}
\tablerefs{1. \citet{Manara2022}; 2. \citet{Sturm2022}; 3. \citet{Salyk2014}; 4. \citet{Kurtovic2018ApJ...869L..44K}; 
5. \citet{Long2022}; 6. \citet{WilliamsBest2014}; 7. \citet{Kama2016};
8. \citet{Guilloteau2012}; 9. \citet{Bergner2019}; 10. \citet{Tripathi2017};
11. \citet{Braun2021}; 12. \citet{Huang2023ApJ...943..107H}; 13. \citet{Long2019};
14. \citet{Fang2018}; 15. \citet{Pascucci2020}; 16. \citet{Kama2016theory}
}
\end{deluxetable*}

\bibliography{lit}{}
\bibliographystyle{aasjournal}



\end{document}